\documentclass{article}

\usepackage[preprint]{neurips_2024}
\usepackage{fancyhdr} 
\usepackage{blindtext} 
\usepackage{makecell}
\usepackage{natbib}

\usepackage[utf8]{inputenc} 
\usepackage[T1]{fontenc}    
\usepackage{hyperref}       
\usepackage{url}            
\usepackage{booktabs}       
\usepackage{amsfonts}       
\usepackage{nicefrac}       
\usepackage{microtype}      
\usepackage{xcolor}         
\usepackage{amsmath}
\usepackage{float}
\usepackage{tabularray}
\usepackage{multirow}
\usepackage{booktabs, caption, array}
\usepackage{amssymb}
\usepackage{lipsum}         
\usepackage{CJKutf8}        
\usepackage{nomencl}
\usepackage{glossaries}
\usepackage{xspace}
\usepackage{multirow}
\usepackage{graphicx}
\usepackage{bbding}
\usepackage{pifont}
\usepackage{algorithm}
\usepackage{subfigure}
\usepackage{enumitem}
\usepackage{multicol}
\usepackage{caption}
\usepackage{array}
\usepackage{fp}
\usepackage{multirow}   
\usepackage{makecell}
\usepackage{flushend}
\usepackage{cases}
\usepackage{color}
\usepackage{algpseudocode}
\usepackage{listings}
\usepackage{mathrsfs}
\usepackage{longtable}
\usepackage{colortbl}
\usepackage{bm}
\usepackage{caption}

\setlist[itemize]{noitemsep, topsep=0pt}
\usepackage{arydshln}
\usepackage{lipsum}
\definecolor{codegreen}{rgb}{0,0.3,0.6}
\definecolor{codegray}{rgb}{0.5,0.5,0.5}

\newcommand{\ie}{\emph{i.e.,}\xspace}
\newcommand{\eg}{\emph{e.g.,}\xspace}

\newcommand{\paratitle}[1]{\vspace{1.5ex}\noindent\textbf{#1}}

\newcommand{\ignore}[1]{}

\usepackage[most]{tcolorbox}
\definecolor{darkorange}{RGB}{255, 140, 0}
\definecolor{lightgreen}{RGB}{145, 204, 117}
\definecolor{lightyellow}{RGB}{250, 200, 88}
\definecolor{lightred}{RGB}{238, 102, 102}
\definecolor{lightblue}{RGB}{115, 192, 222}

\definecolor{gray_1}{HTML}{B7B7B7}
\definecolor{gray_2}{HTML}{F0F0F0} 
\definecolor{frame_blue}{HTML}{A9D18E}

\newtcolorbox[auto counter, number within=section]{PromptBox}[2][]{
    enhanced,
    breakable,
    colback=gray_2, 
    colframe=gray_1,
    coltitle=white,
    fontupper=\small,
    fonttitle=\bfseries,
    title={#2}, 
    label={#1},
    arc=2pt,
    boxrule=1pt,
    left=2mm, right=2mm, top=2mm, bottom=2mm,
}

\usepackage{etoolbox}
\makeatletter
\patchcmd{\@maketitle}
  {\@toptitlebar}
  {%
    \hbox to \textwidth{%
    
    \includegraphics[height=0.7cm]{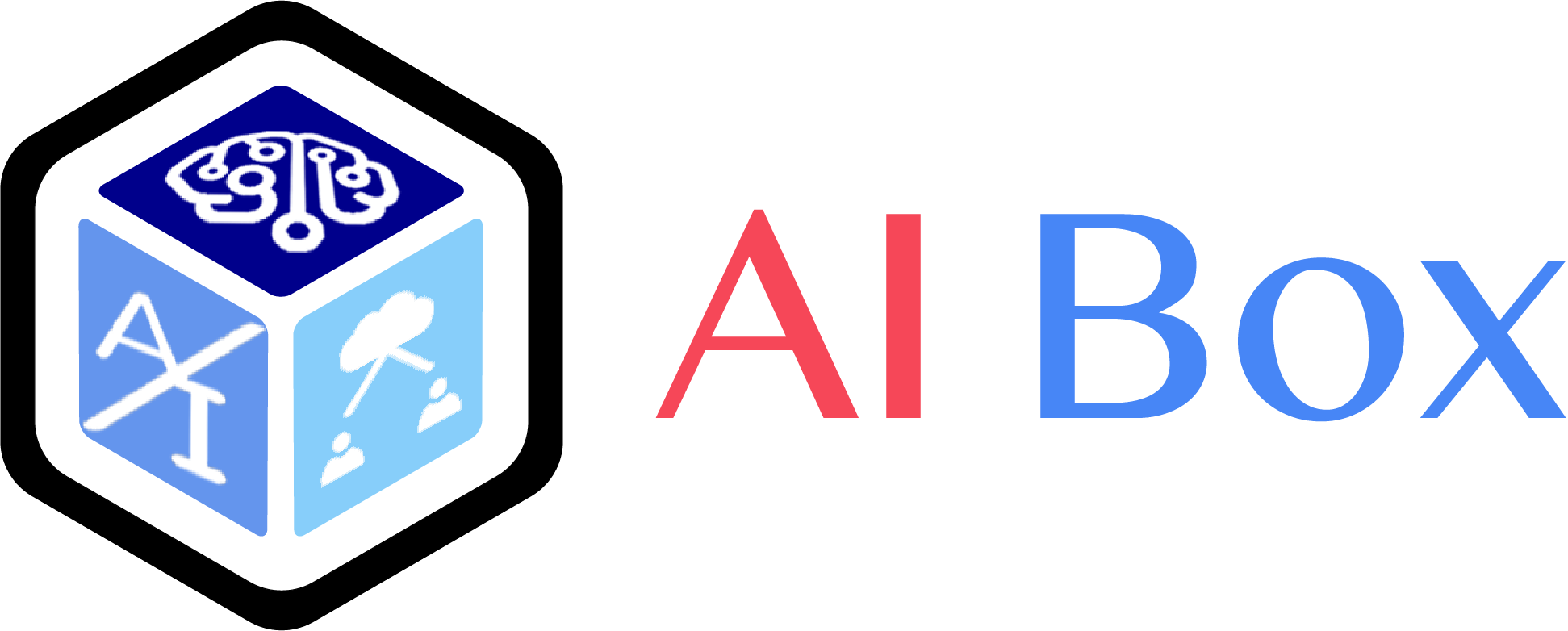}%
      \hfill
      \includegraphics[height=0.7cm]{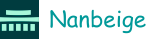}%
    }%
    \vspace{0.3em}
    \@toptitlebar
  }
  {}{}
\makeatother

\title{SWE-World: Building Software Engineering Agents in\\ Docker-Free Environments}

\author{%
   Shuang Sun$^{1}$\thanks{Equal contribution.}~,
  Huatong Song$^{1*}$,
  Lisheng Huang$^{1*}$,
  Jinhao Jiang$^{1*}$,\\
  \textbf{Ran Le$^2$, Zhihao Lv$^{1}$, Zongchao Chen$^{2}$},
  \textbf{Yiwen Hu$^{1}$},
  \textbf{Wenyang Luo$^{1}$},\\
  \textbf{Wayne Xin Zhao$^{1}$\thanks{Correspondence to Wayne Xin Zhao and Yang Song.}~, }
  \textbf{Yang Song$^{2\dagger}$, } 
  \textbf{Hongteng Xu$^{1}$, } 
   \textbf{Tao Zhang}$^{2}$, \textbf{Ji-Rong Wen}$^{1}$
  \\
  $^1$Gaoling School of Artificial Intelligence, Renmin University of China.\\
  $^2$BOSS Zhipin, Beijing, China.\\
  \texttt{sunshuang@ruc.edu.cn}, \texttt{batmanfly@gmail.com}, \texttt{songyang@kanzhun.com}
}

\begin{document}
\maketitle



\definecolor{babyblue}{HTML}{F8F9FE}
\newtcolorbox{bluebox}{
  colback=babyblue,    
  colframe=babyblue,  
  width=1.0\textwidth,  
  center,               
  arc=8pt,                 
  boxrule=0pt,           
  boxsep=0pt,       
  left=2pt,               
  right=2pt,              
  top=10pt,                
  bottom=10pt              
}

\begin{bluebox}
\begin{abstract}
Recent advances in large language models (LLMs) have enabled software engineering agents to tackle complex code modification tasks. Most existing approaches rely on execution feedback from containerized environments, which require dependency-complete setup and physical execution of programs and tests. While effective, this paradigm is resource-intensive and difficult to maintain, substantially complicating agent training and limiting scalability. We propose SWE-World, a Docker-free framework that replaces physical execution environments with a learned surrogate for training and evaluating software engineering agents. SWE-World leverages LLM-based models trained on real agent–environment interaction data to predict intermediate execution outcomes and final test feedback, enabling agents to learn without interacting with physical containerized environments. This design preserves the standard agent–environment interaction loop while eliminating the need for costly environment construction and maintenance during agent optimization and evaluation. Furthermore, because SWE-World can simulate the final evaluation outcomes of candidate trajectories without real submission, it enables selecting the best solution among multiple test-time attempts, thereby facilitating effective test-time scaling (TTS) in software engineering tasks. Experiments on SWE-bench Verified demonstrate that SWE-World raises Qwen2.5-Coder-32B from 6.2\% to 52.0\% via Docker-free SFT, 55.0\% with Docker-free RL, and 68.2\% with further TTS. The code is available at
\url{https://github.com/RUCAIBox/SWE-World}
\end{abstract}
\end{bluebox}

\section{Introduction}
\label{sec:intro}

Recent years have seen rapid progress in software engineering (SWE) agents that leverage large language models (LLMs)~\citep{zhao2023survey}, which are beginning to deliver practical value in real-world settings~\citep{jimenez2024swebench, sweb-verified}. Such agents typically operate in an iterative agent–environment interaction loop driven by physical environment execution feedback~\citep{wang2025openhands, minisweagent}. In SWE scenarios, this environment corresponds to an isolated, dependency-complete workspace instantiated from the target repository, most commonly realized via containerization frameworks such as Docker, where programs and unit tests can be executed reliably~\citep{swe-gym, badertdinov2025swerebench}. Consequently, existing approaches to training and evaluating SWE agents highly depend on physical execution environments, which serve as a central prerequisite for effective agent optimization~\citep{yang2025swesmith}.

In practice, execution feedback in SWE tasks encompasses both intermediate program behaviors (\eg reproducing failures) and running unit tests for final evaluation. In particular, evaluation requires executing the agent-fixed repository and verifying that all designated tests pass. Compared to lightweight execution settings that require only a basic interpreter/runtime and limited dependencies (\eg algorithmic programming tasks), SWE tasks necessitate a full repository workspace together with a dependency-complete environment, requiring environment setup, dependency resolution, and program execution, and thus making execution feedback substantially more resource-intensive to construct and maintain in practice~\cite{yang2025kimi}. To support such execution-based training, prior work has predominantly relied on Docker to instantiate task-specific environments and provide reliable execution feedback for supervised fine-tuning (SFT) or reinforcement learning (RL), enabling agents to improve their performance through multi-turn environment interaction~\citep{tao2026swe, deepswe2025,zeng2026davinci}.

Despite their effectiveness, Docker-based training paradigms introduce several fundamental scalability limitations for SWE agents. At a high level, these limitations manifest across data, training, and test-time regimes. First, data scalability is constrained: many real-world repositories and pull requests cannot be readily leveraged, as complex or brittle dependency configurations often prevent projects from building or executing reliably within containerized environments~\citep{wang2025swe}. Second, training scalability is hindered: the storage, management, and distribution of large numbers of Docker images impose substantial infrastructure overhead, which significantly complicates large-scale optimization, particularly reinforcement learning, in resource-constrained academic settings~\citep{deepswe2025}. Third, test-time scalability is limited: because environment interactions are computationally expensive and often irreversible, it becomes difficult to fully exploit additional test-time computation through iterative exploration or trial-and-error strategies~\citep{jain2025r2e}.

Motivated by recent advances in world modeling with LLMs~\citep{copet2025cwm, luo2025mcp}, we explore an alternative paradigm that replaces physical execution environments with learned models. This raises a central question: \textit{can execution feedback traditionally obtained from Docker-based environments be approximated by LLMs, enabling Docker-free training and deployment of software engineering agents?} If feasible, such an approach would fundamentally alleviate the scalability bottlenecks of Docker-centric SWE pipelines by decoupling agent optimization from costly environment instantiation. However, modeling execution at the repository level is inherently challenging: it requires reasoning over large, evolving codebases where localized edits may induce complex, non-local effects, and demands accurate prediction of diverse execution behaviors across heterogeneous projects. In this work, we address this challenge by training LLMs on real agent–environment interaction data to predict execution outcomes, yielding a learned surrogate environment that provides effective feedback for SWE agent training while overcoming the data, training, and test-time scalability constraints of Docker-based approaches.

To this end, we propose \textbf{SWE-World}, a Docker-free framework that replaces resource-intensive physical execution environments with a learned surrogate for training SWE agents. The design of SWE-World is motivated by a key observation: while dependency-complete, runnable environments required for program and test execution are costly to construct and maintain, many agent actions—such as file navigation, text inspection, and code editing (\eg \texttt{ls}, \texttt{grep}, \texttt{vim})—only involve lightweight file-system operations and incur negligible computational overhead. SWE-World explicitly separates these two classes of actions during the interaction loop. Lightweight operations are handled deterministically by a sandbox that directly updates the repository state, whereas execution-oriented commands that would traditionally require Docker are routed to an LLM-based transition model that predicts execution outcomes without instantiating runnable environments. By learning to approximate execution behavior observed in Docker, the transition model effectively replaces dependency-complete containers as the source of execution feedback, allowing agents to receive feedback without incurring the high resource cost of physical execution. Upon episode termination, \ie when the agent submits its final code patch, an LLM-based reward model replaces containerized test execution by acting as a virtual test runner that evaluates the patch and produces structured test feedback together with a binary success signal. Together, these components form a unified surrogate world for the agent, capturing both step-level dynamics and terminal outcomes, alleviating the scalability bottlenecks of resource-heavy execution.

We conduct extensive experiments to evaluate the effectiveness of \textbf{SWE-World} as a Docker-free training and inference environment. Using a combination of open-source benchmarks and newly crawled tasks, we show that agents optimized entirely without physical execution environments achieve strong performance on SWE-bench Verified~\cite{sweb-verified}. In particular, when trained with \emph{Docker-free execution feedback} provided by SWE-World, Qwen2.5-Coder-32B improves from a base resolve rate of 6.2\% to 52.0\%, and further to 55.0\% with additional Docker-free RL optimization, while the smaller Qwen3-4B-Instruct reaches 25.6\% and 30.0\%, respectively. Beyond training, SWE-World also enables effective \emph{Docker-free test-time scaling}: by using the learned reward model to evaluate and select candidate solutions, Qwen2.5-Coder-32B attains a resolve rate of 68.2\% with TTS@8. These results demonstrate that SWE-World can match, and in some settings surpass the effectiveness of Docker-based pipelines, while drastically lowering the infrastructure and resource barrier of SWE experimentation, enabling large-scale SWE research and iteration without industrial-grade container infrastructure.

Our main contributions are summarized as follows:

$\bullet$ Docker-Free SWE Environment: We propose SWE-World, a Docker-free framework that replaces physical execution environments with a learned surrogate for training and evaluating software engineering agents.

$\bullet$ Effective Agent Training without Execution: We show that LLM-based environment feedback can successfully support SFT and RL, achieving performance comparable to or better than training with real execution.

$\bullet$ Scalable Use of Open-Source SWE Data: By eliminating the requirement for buildable environments, SWE-World enables substantially broader utilization of real-world GitHub data for training software engineering agents.

\section{Related Work}

\paratitle{Datasets for Software Engineering Tasks.} The evaluation paradigm for large language models (LLMs) in software engineering has transitioned from isolated code generation tasks~\citep{chen2021evaluating, jain2024livecodebench} toward complex, repository-level issue resolution. SWE-bench~\citep{jimenez2024swebench} and its refined counterpart, SWE-bench-verified~\citep{sweb-verified}, pioneered this shift by establishing a rigorous benchmark for assessing the capacity of LLM-based code agents to resolve real-world software issues. 
Building upon this, several studies have focused on synthesizing issue-solving tasks from GitHub data to facilitate model training~\citep{swe-gym}. SWE-Gym~\citep{swe-gym} and SWE-rebench~\citep{badertdinov2025swerebench} directly harvest interactive SWE tasks from diverse GitHub repositories, while SWE-smith~\citep{yang2025swesmith} introduces an automated pipeline for large-scale issue generation. However, a significant bottleneck in these existing datasets is their heavy infrastructure dependency; each data sample typically necessitates a dedicated Docker environment for execution and verification, incurring substantial computational and storage overhead~\citep{yang2025kimi}. In contrast, our approach streamlines this process by directly filtering issues from GitHub and utilizing only the underlying code repositories and unit tests for trajectory synthesis. By bypassing the requirement for per-sample containerization, our method significantly reduces the resource barrier while maintaining high fidelity in training data.

\paratitle{Software Engineering LLMs and Agents.} To unlock the potential of LLMs as autonomous software agents, researchers have introduced various agentic frameworks, such as SWE-agent~\citep{yang2024swe} and OpenHands~\citep{wang2025openhands}, which provide the necessary infrastructure for environment interaction. Current research in autonomous issue resolution generally follows two paradigms. The first is the agent-based approach, which situates models within authentic, sandboxed environments (\eg Docker) to generate interaction trajectories and perform agentic SFT~\citep{sun2025simpledeepsearcherdeepinformationseeking,tao2026swe,wang2025swe, liu2025context} and RL~\citep{cao2025skyrl,deepswe2025,song2025r1,golubev2025training} training. While effective, this paradigm suffers from significant resource overhead, as each data sample typically demands a dedicated container. The second is the agentless approach~\citep{xia2024agentless, yang2025kimi}, which decomposes software engineering tasks into a structured, three-stage pipeline: fault localization, code repair, and patch verification~\citep{xie2025swe, SWESwiss2025}. Although more efficient, these predefined workflows often restrict the agent's capacity for autonomous exploration and adaptive reasoning. Diverging from these two paths, our method maintains an agentic loop to preserve exploratory autonomy but introduces an LLM-based environment simulator. By simulating environmental feedback rather than relying on heavy-weight containerization, our approach eliminates the infrastructure bottleneck of traditional agent-based systems while maintaining the flexibility of autonomous agents.

\section{Preliminaries}
\label{sec:prelim}

\begin{figure}[t]
    \centering
    \includegraphics[width=1.0\linewidth]{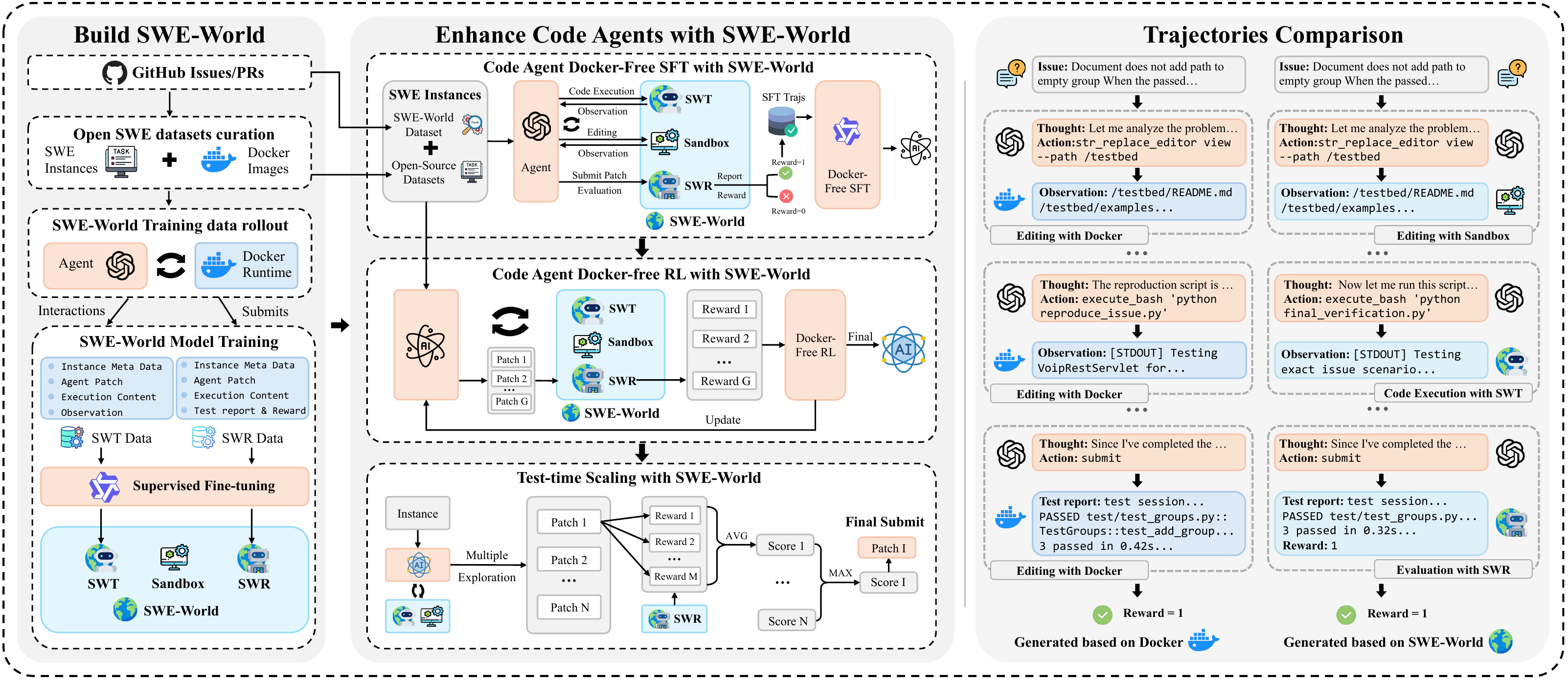}
    \caption{
        Overview of \textbf{SWE-World}.
        \textbf{Left:} We collect agent-Docker interaction data to train the SWE-World Transition Model (SWT) and SWE-World Reward Model (SWR).
        \textbf{Middle:} SWE-World forms a Docker-free surrogate environment, enabling scalable agent enhancement via SFT, RL, and Test-Time Scaling.
        \textbf{Right:} Comparison of Code Agent trajectories generated based on Docker and SWE-World.
    }
    \label{fig:main}
\end{figure}

We study software engineering (SWE) issue resolution as an interactive process involving three components:
(i) a \emph{code agent} that proposes edits and execution commands,
(ii) a \emph{codebase} (repository workspace) to be modified, and
(iii) an \emph{execution environment} $\mathcal{E}$ that provides step-level feedback and evaluates candidate fixes.

\paratitle{SWE Task Instances.}
A task instance is defined by an issue description and a reference repository snapshot.
We denote an instance as $\mathcal{I}=(R,b,d,\mathcal{U})$, where $R$ is a repository, $b$ is a base commit, $d$ is the problem statement (\eg a bug report), and $\mathcal{U}$ is the validation unit tests that must pass for a correct fix.
Additional instance metadata is provided in Appendix~\ref{app:instance_format}.

\paratitle{Interactive Repair Process.}
The agent operates in a mutable workspace initialized at $(R,b)$ and interacts with an execution environment $\mathcal{E}$.
At each step $t$, the agent produces a thought $z_t$ and takes an action $a_t$; the environment $\mathcal{E}$ executes $a_t$ and returns step-level feedback $y_t$:
\begin{equation}
    y_t \sim \mathcal{E}(a_t;\, R,b,P_{t}).
\end{equation}
We represent code edits as a patch $P$ (a set of line-level modifications to the codebase).
A repair episode terminates when the agent takes a \texttt{submit} action, producing the final patch $P$.
A complete trajectory is denoted as
\begin{equation}
\tau = \big[\mathcal{I}, (z_1,a_1,y_1), \ldots, (z_T,a_T,y_T), P\big].
\end{equation}

\paratitle{Evaluation.}
Given a submitted patch $P$, the environment $\mathcal{E}$ evaluates the modified workspace by running the designated unit tests $\mathcal{U}$.
This yields a final evaluation output
\begin{equation}
    y_{\text{eval}} = \langle \texttt{test\_report}, r \rangle \sim \mathcal{E}(P; R,b,\mathcal{U}),
\end{equation}
where \texttt{test\_report} summarizes execution logs, and the binary reward $r\in\{0,1\}$ follows the standard criterion:
\begin{equation}
    r = 1 \;\Longleftrightarrow\; \text{all required unit tests in } \mathcal{U} \text{ pass under } P,
\end{equation}
and $r=0$ otherwise.

\subsection{Execution Environments for SWE}
\label{subsec:env_prelim}

A central challenge in SWE is the \emph{execution environment} $\mathcal{E}$: it must support both lightweight interactions (file navigation and editing) and heavyweight repository-specific execution (running programs and unit tests)~\citep{yang2025swesmith}.
To make this clear, we distinguish four increasingly capable layers of environment that often appear implicitly in prior work.

\paratitle{File System.}
The repository workspace---files and directories that the agent reads and modifies---is the mutable state to transform from $(R,b)$ into a correct fix.

\paratitle{Terminal.}
A terminal provides generic interfaces (\eg \texttt{ls}, \texttt{grep}, \texttt{vim}) and editing tools to navigate and modify the file system, but does not inherently provide repository-specific code execution.

\paratitle{Sandbox.}
A sandbox couples the file system with a terminal to deterministically support navigation/editing for any repository. It typically cannot run repo-specific programs or tests due to missing dependencies and runtime setup.

\paratitle{Docker.}
Docker instantiates a dependency-complete environment for a given repository snapshot, subsuming sandbox operations while enabling repo-specific code execution and test runs.
However, building and maintaining Docker images is brittle in practice: different repositories require different installation procedures and dependency constraints, and large-scale SWE training requires spawning and managing many images and containers.
As a result, the storage, management, and distribution of Docker impose significant infrastructure overhead.

\subsection{From Docker-Based to LLM-Based Environments}
\label{subsec:llm_env_prelim}

The above decomposition suggests a key observation:
many agent actions in SWE only require the lightweight sandbox, while the main scalability bottleneck comes from repository-specific code execution traditionally provided by Docker.
This motivates our core idea: replace the Docker execution component with learned models, while retaining a deterministic sandbox for file operations.

Concretely, we aim to construct a surrogate environment $\mathcal{E}_{\text{world}}$ that preserves the standard agent--environment interface:
the sandbox maintains the workspace state and supports navigation/editing actions,
while LLMs approximate the execution feedback and test-based evaluation that would otherwise require Docker.
By combining a universal sandbox with LLM-based execution surrogates, we can eliminate the need for Docker-based environments during training and inference, substantially improving scalability.
Details of the learned components are introduced in Section~\ref{sec:swe_world}.

\section{SWE-World: LLM-Based Docker-Free Environment}
\label{sec:swe_world}

We propose \textbf{SWE-World}, a surrogate execution environment $\mathcal{E}_{\text{world}}$ that approximates containerized runtimes by training LLMs to predict execution feedback from real agent--environment interaction traces.
By decoupling the agent from physical execution, SWE-World enables scalable training and inference without Docker.

\subsection{System Architecture}
\label{subsec:system}

SWE-World replaces Docker with a universal lightweight sandbox and learned LLM components.
Given an agent action $a_t$, we categorize it into two types:
\begin{itemize}[leftmargin=2.0em]
    \item Navigation \& Editing: file exploration and edits via generic shell/tool commands (\eg \texttt{ls}, \texttt{cat}, \texttt{grep}, \texttt{view}, \texttt{create}, \texttt{str\_replace}).
    \item Code Execution: repository-specific execution commands whose outputs depend on runtime semantics (\eg \texttt{python reproduce.py}, \texttt{pytest}).
\end{itemize}

Navigation and editing actions are executed deterministically by a lightweight \textbf{Sandbox}, which maintains the workspace state and returns step feedback $y_t$.
Code execution actions are handled by the learned \textbf{SWE-World Transition Model} (SWT), which predicts step-level execution feedback $\hat{y}_t$.
When the agent terminates by submitting a final patch $P$, the \textbf{SWE-World Reward Model} (SWR) acts as a virtual test runner and produces a test report together with the binary reward.
To drive the learned models, we construct a compact \emph{context} $\kappa$ that includes the instance information and the current workspace state; concrete context definitions are given below.

\subsubsection{Lightweight Sandbox for Navigation and Editing}
\label{subsubsec:sandbox}

The sandbox consists of the file system and terminal: it deterministically supports navigation and editing over the repository workspace.
We delegate these operations to the sandbox (instead of an LLM) for two reasons:
(1) Reliability: file operations are deterministic; LLM simulation may hallucinate files or contents and catastrophically mislead the agent.
(2) Efficiency: navigation/editing is inexpensive and does not require semantic reasoning, making LLM inference unnecessary.
As a result, the workspace state and the code content referenced by the learned models remain accurate and strictly consistent with the agent's edit history.

\subsubsection{SWT: Step-Level Execution Feedback Simulation}
\label{subsubsec:swt}

For Code Execution actions, we employ the SWE-World Transition Model, denoted as $\mathcal{M}_{\text{SWT}}$.
Modeling execution at the repository level is inherently challenging: it requires reasoning over large, evolving codebases where localized edits may induce complex, non-local effects, and demands accurate prediction of diverse execution behaviors across heterogeneous projects.
Moreover, SWT must be sensitive to the agent's incremental edits, producing meaningfully different feedback as the patch evolves.

\paratitle{Transition Context.}
At step $t$, we construct an transition context $\kappa^{\text{SWT}}_t$ consisting of three parts:
\begin{itemize}[leftmargin=2.0em]
    \item Instance Meta Data: problem description, an initial analysis generated by a powerful LLM (summarizing the core bug and intended fix), and a ground-truth patch as an internal reference. This reference is strictly hidden from the agent to prevent leakage.
    \item Agent Patch: the current patch reflecting the agent's modifications so far.
    \item Execution Content: the command~(action $a_t$) to be simulated and the relevant code content needed to determine the runtime behavior.
\end{itemize}

Given an context $\kappa^{\text{SWT}}_t$, SWT predicts step-level execution feedback:
\begin{equation}
    \hat{y}_t = \langle \texttt{stdout}, \texttt{stderr}, \texttt{exit\_code} \rangle \sim \mathcal{M}_{\text{SWT}}(\kappa^{\text{SWT}}_t).
\end{equation}
By conditioning on $\kappa^{\text{SWT}}_t$, SWT can produce realistic feedback such as error tracebacks and printed logs, enabling iterative debugging and refinement without Docker execution. See Appendix~\ref{app:instance_format} for transition context details.

\subsubsection{SWR: Test Report and Reward Generation for Evaluation Simulation}
\label{subsubsec:swr}

The SWE-World Reward Model, denoted as $\mathcal{M}_{\text{SWR}}$, validates the agent's final submission.
Prevalent approaches cast validation as a generative classification task, prompting an LLM to map the agent trajectory directly to a binary token (\ie \textsc{Yes}/\textsc{No})~\cite{shum2025swe, tao2026swe}.
During inference, they extract token log-probabilities $l_{\textsc{yes}}$ and $l_{\textsc{no}}$ and compute a scalar score via softmax:
\begin{equation}
    \hat{r} = \frac{\exp(l_{\textsc{yes}})}{\exp(l_{\textsc{yes}}) + \exp(l_{\textsc{no}})}.
\end{equation}
This black-box formulation provides limited interpretability and can be sensitive to long, noisy trajectories.

In contrast, SWR acts as a \emph{virtual test runner}, simulating the execution of the unit tests $\mathcal{U}$ on the final submitted patch $P$ and generating a structured test report before assigning the final binary reward.
Moreover, SWR must faithfully account for complex test logic and aggregate outcomes across many test cases, where a single failing case determines $\hat{r}=0$.

\paratitle{Evaluation Context.}
For a trajectory $\tau$, to drive this simulation, we construct an evaluation context $\kappa^{\text{SWR}}_{\tau}$.
It contains the same three parts as the transition context (Instance Meta Data, Agent Patch, and Execution Content), and additionally includes the unit tests $\mathcal{U}$:
\begin{itemize}[leftmargin=2.0em]
    \item Fail to Pass (F2P): tests that reproduce the issue and must transition to passing.
    \item Pass to Pass (P2P): regression tests that must remain passing.
\end{itemize}
In particular, for $\kappa^{\text{SWR}}_{\tau}$, the Agent Patch is the final submitted patch $P$, and the Execution Content corresponds to the test command and the test-case content of $\mathcal{U}$.

Given $\kappa^{\text{SWR}}_{\tau}$, SWR predicts the evaluation output:
\begin{equation}
    \hat{y}_{\text{eval}} = \langle \texttt{test\_report}, \hat{r} \rangle \sim \mathcal{M}_{\text{SWR}}(\kappa^{\text{SWR}}_{\tau}), \qquad \hat{r}\in\{0,1\}.
\end{equation}
By enforcing the generation of a detailed test report prior to $\hat{r}$, SWR provides an interpretable verification signal that supports scalable Docker-free selection and optimization. See Appendix~\ref{app:instance_format} for evaluation context details.

\subsection{Training SWT and SWR}
\label{subsec:sw_training}

We train SWT and SWR based on Qwen2.5-Instruct-32B and Qwen2.5-Instruct-72B~\citep{qwen2025qwen25technicalreport} via SFT, utilizing data collected directly from the interactions between SWE agents and real execution environments.

\subsubsection{Data Collection from Real Docker Rollouts}

We construct our training corpus by generating trajectory rollouts on open-source SWE datasets~\citep{jain2025r2e, swe-gym, badertdinov2025swerebench} and corresponding Docker images.
For each code execution step, we extract $\kappa^{\text{SWT}}_t$ and the real execution feedback $y_t$ directly from the Docker container's output, yielding transition samples:
\begin{equation}
    \mathcal{D}_{\text{trans}} = \{ (\kappa^{\text{SWT}}_t, y_t) \}.
\end{equation}
For a trajectory $\tau$, at the terminal step, we extract $\kappa^{\text{SWR}}_{\tau}$ execute the unit tests to obtain the evaluation outcome $y_{\text{eval}}$, forming the reward dataset:
\begin{equation}
    \mathcal{D}_{\text{reward}} = \{ (\kappa^{\text{SWR}}_{\tau}, y_{\text{eval}}) \}.
\end{equation}
Crucially, this collection process achieves high data efficiency: a single rollout simultaneously yields multiple interaction samples for supervising the environment models and an agent trajectory for policy training. Details are provided in Appendix~\ref{app:training_details}.

\subsubsection{Reverse-Reasoning Distillation for CoT Backfilling}

Directly mapping complex repository contexts to precise execution feedback poses a significant learning challenge. Chain-of-Thought (CoT)~\citep{wei2022chain} reasoning mitigates this by providing intermediate logical steps, effectively breaking down the reasoning process and enhancing model learnability. Moreover, during inference, the explicit generation of CoT facilitates more granular analysis, leading to higher prediction accuracy~\cite{huang2025transformers,kojima2022large}.

To obtain high-quality CoT data, we employ a reverse-reasoning strategy~\citep{chen2025reverse,lin2025scaling}. We feed both the input context $\kappa$ and the ground-truth execution feedback $y_{\text{GT}}$ (from Docker) to a powerful reasoning model, requiring it to produce a strictly forward derivation without explicitly revealing the answer in the reasoning process. This model is prompted to generate a step-by-step derivation CoT that reasons from the input to the specified output:
\begin{equation}
    \text{CoT} \leftarrow \pi_{\text{teacher}}(\text{Reasoning} \mid \kappa, y_{\text{GT}}).
\end{equation}
Before integration, we employ an LLM-as-a-Judge~\citep{gu2024survey} to assess the quality of the generated CoTs, filtering out invalid reasoning, or CoTs that directly leak $y_{\text{GT}}$.
We backfill the high-quality thoughts into our dataset, formatting the target output by wrapping the thoughts within \texttt{<think>} tags followed immediately by the ground truth. 
Consequently, both SWT and SWR are trained to predict the sequence $\texttt{<think>}\text{CoT}\texttt{</think>}y_{\text{GT}}$, enabling them to learn the underlying execution logic more effectively.
 Details are provided in Appendix~\ref{app:training_details}.


\section{Training SWE Agents with SWE-World}

In this section, we leverage \textbf{SWE-World} to establish an end-to-end, fully Docker-free training pipeline for training SWE agents, enabling scalable data curation, supervised fine-tuning, and reinforcement learning.

\subsection{Software Eengineering Data Preparation}

We construct a unified instance pool from two sources.

\paratitle{Open-Source SWE Datasets.} We aggregate instances from open-source SWE datasets, including R2E-Gym, SWE-Gym, and SWE-rebench. We convert these instances into a unified schema compatible with SWE-World.

\paratitle{SWE-World Dataset.} A limitation of previous dataset construction pipelines is the heavy reliance on Docker environments; a vast number of Pull Requests (PRs) and issues are typically discarded simply because their repositories fail to build in a container~\citep{yang2025swesmith}.
Leveraging SWE-World's ability to simulate execution without physical constraints, we bypass this bottleneck. We crawled a fresh collection of PRs and issues from GitHub, applied deduplication against existing datasets, and parsed the relevant repository information.
After applying heuristic filtering rules, we obtain \textbf{SWE-World Dataset}, a curated set of 16.6K high-quality instances. As shown in Table~\ref{tab:dataset_comparison_main}, the SWE-World Dataset covers more tasks and far more repositories than prior SWE datasets, offering broader coverage and higher diversity. The detailed data processing pipeline is provided in the Appendix~\ref{app:dataset_stats}. The final training set is a mixture of these open-source and newly curated instances.

\begin{table}[h]
    \small
    \centering
    \caption{Comparison of SWE-World Dataset and other SWE datasets with Docker environments.}
    \label{tab:dataset_comparison_main}
    \setlength{\tabcolsep}{10pt}
    \renewcommand{\arraystretch}{1.2}
    \begin{tabular}{l c c c}
        \toprule
        \textbf{Dataset} & \textbf{\# Tasks} & \textbf{\# Repos} & \textbf{Source} \\
        \midrule
        SWE-rebench & 6.5k & 1,429 & Real  \\
        SWE-Gym & 2.4k & 11 & Real \\
        R2E-Gym & 4.6k & 10 & Synth \\
        \midrule
        \textbf{SWE-World Dataset} & \textbf{16.6k} & \textbf{3,763} & \textbf{Real} \\
        \bottomrule
    \end{tabular}
\end{table}

\subsection{Docker-Free Supervised Fine-tuning}

We employ a rejection sampling fine-tuning strategy to train the policy model. This process consists of three key steps:

\paratitle{Trajectory Generation.} For each training instance, we use a powerful code agent~\citep{glm_46, minimax_m2}  to interact within the SWE-World environment (including the SWT and the Sandbox). This interaction yields agent trajectories efficiently and in parallel, bypassing the overhead of container management.

\paratitle{Dual-Stage Filtering.} To ensure the quality of the training data, we apply a rigorous filtering mechanism to the generated agent trajectories:
\begin{itemize}[leftmargin=2.0em]
    \item \textit{Rule-Based Filtering:} We discard agent trajectories exhibiting invalid tool usage, execution timeouts, excessive context length, exceeding the maximum turn limit, or failure to produce a final submission.
    \item \textit{SWR-Based Verification:} We use the SWR to assess the correctness of the remaining agent trajectories. We retain only those receiving a positive reward ($\hat{r}=1$), which indicates that the agent has successfully resolved the issue according to the surrogate evaluation.
\end{itemize}

\paratitle{Agentic Supervised Fine-Tuning.} Finally, we use the high-quality agent trajectories to train our policy model~(Qwen2.5-32B-Coder-Instruct~\citep{hui2024qwen2} and Qwen3-4B-Instruct-2507~\citep{yang2025qwen3technicalreport}). Following standard practices for agentic training, we perform agentic SFT~\citep{wang2025swe}  on the agent's thoughts and actions. Crucially, the entire SFT pipeline is \textit{Docker-free}.
Training details are provided in Appendix~\ref{app:ef_sft_details}.

\subsection{Docker-Free Reinforcement Learning}
\label{subsec:docker_free_rl}

SWE reinforcement learning~\citep{golubev2025training, cao2025skyrl}  is expensive and brittle due to its heavy reliance on Docker containers, since repeatedly spawning and managing containers for rollouts and validation consumes massive memory, CPU, and storage resources and often destabilizes the training infrastructure.
SWE-RM~\citep{shum2025swe}  replaces part of the validation with a learned verifier, but still relies on Docker containers for transition feedback during rollouts and mixes execution-based signals for final rewards, leaving Docker in the RL loop. In contrast, building on SWE-World, we achieve fully Docker-free RL.

\paratitle{Agentic RL Loop with SWE-World.}
Starting from the SFT-trained model, we further optimize the agent with RL. We first launch SWT and SWR as inference services. During training rollouts, the agent interacts with SWE-World, receiving transition feedback from the SWT and Sandbox, while the SWR assigns the final trajectory reward. Crucially, this pipeline eliminates the need for Docker containers, requiring only the maintenance of inference servers for SWT and SWR models.

\paratitle{Optimization.}
We optimize the policy following the Group Relative Policy Optimization (GRPO) paradigm~\citep{shao2024deepseekmath,deepswe2025}. Specifically, we employ a stabilized variant that incorporates clipped ratio objectives, leave-one-out advantage estimation, and length normalization to effectively handle long-horizon reasoning tasks.

By using SWT for transition feedback and SWR for terminal rewards, SWE-World supports stable, scalable, fully Docker-free RL for SWE agents. Training details are provided in Appendix~\ref{app:ef_rl_details}.

\subsection{Test-Time Scaling with SWR}

We further improve inference performance via test-time scaling (TTS), where multiple candidate solutions are generated and a verifier selects the best one. 
While prevalent verifiers rely on the single-token generative classification discussed previously~\citep{jain2025r2e,swe-gym}, SWR enables verification grounded in simulated execution mechanics.

SWR is explicitly designed to judge whether an agent trajectory has successfully resolved the issue, making it inherently suitable for TTS. Given that SWR outputs a discrete binary reward, we employ a multi-sample voting strategy to derive a fine-grained confidence score.
Specifically, for a given instance, we sample $N$ candidate trajectories using the code agent. For each trajectory $\tau$, we query SWR $M$ times to mitigate variance and calculate the mean reward:

\begin{equation}
    \text{Score}(\tau) = \frac{1}{M} \sum_{i=1}^{M} \hat{r}_i.
\end{equation}
where $\langle \texttt{test\_report}, \hat{r}_i \rangle \sim \mathcal{M}_{\text{reward}}(\kappa^{\text{SWR}}_{\tau})$ and $\kappa^{\text{SWR}}_{\tau}$ denotes the evaluation context derived from the candidate trajectory $\tau$.
The trajectory with the highest average score is selected as the final submission.

\section{Experiments}

In this section, we first detail the training implementation for the SWE-World components (SWT and SWR) and SWE agents. We then present our evaluation results, ablation studies, and further analyses to validate our motivation and methodology.

\subsection{Experimental Settings}
\label{subsec:exp_settings}

\paratitle{Agent Scaffolding.}
We build our agent on top of R2E-Gym framework~\citep{jain2025r2e}, which is a minimal scaffold adapted from OpenHands~\citep{wang2025openhands} and follows a standard ReAct-style interaction loop~\citep{yao2022react}.
The agent is equipped with three tools: \texttt{str\_replace\_editor} for file reading and editing, \texttt{execute\_bash} for running shell commands, and \texttt{submit} tool for terminating the episode.
We integrate SWE-World into the R2E-Gym framework execution backend, enabling the same scaffold to run either with Docker or with SWE-World as the environment.

\paratitle{Evaluation Benchmark.}
We evaluate on SWE-bench Verified~\citep{jimenez2024swebench}, a curated split of 500 real-world GitHub issue--PR tasks over 12 Python repositories.
Performance is measured by \emph{resolve rate} (\%), \ie the fraction of instances whose final patch passes all designated tests in the evaluation harness.

\paratitle{SWT Evaluation.}
We evaluate SWT in an end-to-end setting by replacing Docker-based step feedback with SWT during rollouts, while keeping the code agent fixed (Minimax-M2.1~\citep{minimax_m21}).
After each rollout terminates with a submitted patch, we run the unit test in Docker to obtain the reward and report the resolve rate on SWE-bench Verified.

\paratitle{SWR Evaluation.}
For SWR evaluation, we compute the Accuracy, Precision, Recall, and F1 by comparing the SWR-predicted rewards against the ground-truth rewards obtained via Docker-based test execution on a held-out set of SWE-bench Verified trajectories.

\paratitle{SWE Agents Evaluation.}
We evaluate SWE-World models on the SWE-bench Verified dataset, utilizing a Docker execution backend for verification and benchmarking against the open-source code agents~\citep{hui2024qwen2, yang2025qwen3technicalreport, swe-gym, jain2025r2e, zeng2025skywork, yang2025swesmith, cao2025skyrl,  deepswe2025, wang2025swe, yang2025kimi, sonwane2025bugpilot, tao2026swe, xie2025swe, ma2024lingma, wei2025swe}.
To prevent git hacking~\citep{xiao2026mimo}, we disallow solution-revealing git commands (\eg \texttt{git log}, \texttt{git show}).

\paratitle{Test-Time Scaling Setup.}
When using SWR for test-time scaling, we sample $N{=}8$ candidate trajectories per instance and query SWR $M{=}3$ times per candidate to estimate its expected reward; we report this setting as \textsc{TTS@8}.

\subsection{Main Results}

\begin{table*}[htbp]
    \small
    \centering
    \setlength{\tabcolsep}{6pt}
    \begin{tabular}{lclcc}
        \toprule
        \textbf{Model/Method} & \textbf{Scaffold} & \textbf{Training} & \textbf{Environment} & \textbf{Resolve Rate (\%)} \\
        \midrule
        Qwen2.5-Coder-32B  & OpenHands & - & Docker & 6.2 \\
        Qwen3-32B & OpenHands & - & Docker & 23.2 \\
        Qwen3-Coder-30B-A3B& OpenHands & - & Docker & 51.6 \\
        SWE-Gym-32B& OpenHands & SFT & Docker & 20.6 \\
        R2E-Gym-32B & R2E-Gym & SFT & Docker & 34.4 \\
        \quad + TTS@16 & R2E-Gym & SFT & Docker & 49.4 \\
        Skywork-SWE-32B & OpenHands & SFT & Docker & 38.0 \\
        \quad + TTS@8 & OpenHands & SFT & Docker & 47.0 \\
        SWE-agent-LM-32B& SWE-agent & SFT & Docker & 40.2 \\
        SWE-Fixer-72B  &  Agentless  & SFT & - & 32.8 \\
        SA-SWE-32B  & OpenHands & RL & Docker & 39.4 \\
        Llama3-SWE-RL-70B&  Agentless & SFT+RL & - & 41.0 \\
        Lingma-SWE-GPT-72B&  Agentless & SFT & - & 30.2\\
        DeepSWE-32B-Preview & OpenHands & RL &  Docker & 42.2 \\
        \quad + TTS@16 & OpenHands & RL & Docker & 59.0 \\
         Kimi-Dev-72B& SWE-Agent   & SFT+RL &- & 48.6 \\
        \quad + TTS@40 & Agentless & SFT+RL & - & 60.4 \\
        SWE-Mirror-LM-32B & MOpenHands & SFT & Docker & 52.2 \\
        FrogBoss-32B& SWE-Agent  & SFT+RL  & Docker  & \underline{54.6} \\
        SWE-Lego-Qwen3-32B& OpenHands & SFT & Docker & 52.6 \\
        \quad + TTS@16& OpenHands & SFT & Docker & 58.8 \\
        \midrule
        \textbf{SWE-World-4B-SFT}& R2E-Gym & SFT & 
\multirow{5}{*}{\makecell{Sandbox \\ + LLMs}} & 25.6 \\
        \textbf{SWE-World-4B-RL}& R2E-Gym & SFT+RL &  & 30.0 \\
        \textbf{SWE-World-32B-SFT}& R2E-Gym & SFT &  & 52.0  \\
        \textbf{SWE-World-32B-RL} & R2E-Gym & SFT+RL &  & \textbf{55.0} \\
        \quad + TTS@8& R2E-Gym & SFT+RL & & 68.2 \\
        \bottomrule
    \end{tabular}
\caption{Performance of various models on SWE-Bench Verified. The best results are in \textbf{bold} and the second-best are \underline{underlined}, excluding test-time scaling (TTS) results.}
\label{tab:main_results} 
\end{table*}

\begin{table}[htbp]
    \small
    \centering

    \begin{minipage}{0.45\textwidth}
        \centering
        \resizebox{\linewidth}{!}{
        \begin{tabular}{l c l}
            \toprule
            \textbf{Transition feedback} & \textbf{Resolve Rate (\%)} & \textbf{$\Delta$} \\
            \midrule
            Docker (GT) & 68.4 & -- \\
            \midrule
            Minmax-M2.1 & 56.2 & $\downarrow 12.2\%$ \\
            GLM-4.7 & 59.4 & $\downarrow 9.0\%$ \\
            \midrule
            SWT-32B & 55.2 & $\downarrow 13.2\%$ \\
            SWT-72B & \textbf{60.2} & $\downarrow \textbf{8.2\%}$ \\
            \bottomrule
        \end{tabular}
        }
        \caption{SWE-bench Verified performance using different transition-feedback providers.}
        \label{tab:main_trans}
    \end{minipage}
    \hfill
    \begin{minipage}{0.45\textwidth}
        \centering
        \resizebox{\linewidth}{!}{
        \begin{tabular}{l c c c c}
            \toprule
            \textbf{Reward Simulation} & \textbf{Acc.} & \textbf{Prec.} & \textbf{Recall} & \textbf{F1} \\
            \midrule
            Docker (GT) & 1.00 & 1.00 & 1.00 & 1.00 \\
            \midrule
            Minmax-M2.1 & 0.740 & 0.709 & \textbf{0.891} & 0.790 \\
            GLM-4.7 & 0.768 & 0.763 & 0.836 &\textbf{0.798} \\
            \midrule
            SWR-32B & 0.754 & 0.779 & 0.770 & 0.774 \\
            SWR-72B & \textbf{0.770} & \textbf{0.780} & 0.807 & 0.794 \\
            \bottomrule
        \end{tabular}
        }
        \caption{Performance of reward simulation against Docker ground truth.}
        \label{tab:main_reward}
    \end{minipage}
\end{table}

\paratitle{Performance of SWE Agents.}
As shown in Table~\ref{tab:main_results}, our method reaches the frontier performance among open-source 32B models.
Starting from the Qwen2.5-Coder backbone (6.2\%), SFT and RL boost performance to 52.0\% and 54.8\% respectively.
Crucially, our purely Docker-free models significantly outperform baselines trained with real Docker execution (\eg FrogBoss-32B at 54.6\%), validating the efficacy of our pipeline.
With SWR-based TTS@8, performance peaks at 68.2\%, surpassing the previous best (Kimi-Dev-72B TTS@40) by a large margin.

\paratitle{Effectiveness of SWT.}
While a performance gap between simulation and reality is inevitable, SWT-72B minimizes this degradation most effectively.
It supports a resolve rate of 60.2\% for Minmax M2.1 (Table~\ref{tab:main_trans}), outperforming general-purpose LLMs like GLM-4.7 (59.4\%) and Minimax-M2.1 (56.2\%) as a more faithful environmental surrogate.

\paratitle{Effectiveness of SWR.}
Table~\ref{tab:main_reward} validates the robustness of our reward models.
SWR-32B achieves an accuracy of 0.754, surpassing Minmax-M2.1, and notably outperforms both Minmax-M2.1 and GLM-4.7 in Precision (0.779).
Scaling up, SWR-72B further advances performance, overtaking GLM-4.7 in both Accuracy (0.770) and Precision (0.780), thereby demonstrating the strongest comprehensive capability in reward simulation.

\subsection{Ablation Study}

\begin{table}[t]
    \small
    \centering
    \begin{tabular}{lcc}
        \toprule
        \textbf{SFT data source} &
        \textbf{Total \#Traj} &
        \textbf{Resolve Rate (\%)} \\
        \midrule
        Docker~(baseline)
            &  5.7K
            &  51.4 \\
        SWE-World
            &  5.7K
            &  52.2 \\
        SWE-World + Docker
            &  9.3K
            &  \textbf{53.8} \\
        \bottomrule
    \end{tabular}
    \caption{
    SFT performance comparison using trajectories collected from Docker, SWE-World, and their mixture, under identical training settings.
}
    \label{tab:sft_mixture_ablation}
\end{table}

We investigate whether replacing the ground-truth Docker environment with SWE-World affects the quality of training data.
We generated two parallel datasets of 5.7K trajectories using identical expert agents and prompt settings, differing only in the execution backend used during the rollout phase.
As shown in Table~\ref{tab:sft_mixture_ablation}, the model fine-tuned on SWE-World trajectories achieves a resolve rate of 52.2\%, slightly outperforming the model trained on Docker-generated data (51.4\%).
Moreover, starting from the 5.7K SWE-World trajectories, we deduplicate an additional Docker-collected set and retain 3.6K unique trajectories; mixing them yields 9.3K trajectories in total and further improves performance to 53.8\%.
This result suggests that our simulated environment provides supervision quality comparable to, or even better than, the physical environment. Furthermore, mixing in Docker-based trajectories yields additional gains, confirming that SWE-World is a highly effective, scalable alternative for data generation.

\section{Further Analysis}
This section provides further analyses on the impact of CoT (Section~\ref{sec:cot_impact}), RL training dynamics (Section~\ref{subsec:rl_dynamics}), test-time scaling behavior (Section~\ref{sec:ana_tts}), and qualitative fidelity of SWT/SWR simulation (Section~\ref{subsec:qual_fidelity}), to clarify the key factors behind SWE-World’s effectiveness.

\subsection{Impact of Chain-of-Thought}
\label{sec:cot_impact}

\begin{table}[htbp]
    \centering
    \setlength{\tabcolsep}{3pt} 

    \begin{minipage}{0.43\textwidth}
        \centering
        \resizebox{\linewidth}{!}{
            \begin{tabular}{lcc}
                \toprule
                \textbf{Method} & \textbf{Data Size} & \textbf{Resolve Rate (\%)} \\
                \midrule
                w/o CoT & 26K & 55.2 \\
                w. CoT  & 26K & \textbf{56.0} \\
                \bottomrule
            \end{tabular}
        }
        \caption{CoT yields only a marginal improvement for SWT.}
        \label{tab:cot_swt}
    \end{minipage}
    \hfill 
    \begin{minipage}{0.52\textwidth}
        \centering
        \resizebox{\linewidth}{!}{
            \begin{tabular}{lccccc}
                \toprule
                \textbf{Method} & \textbf{Data Size} & \textbf{Acc.} & \textbf{Prec.} & \textbf{Recall} & \textbf{F1} \\
                \midrule
                w/o CoT & 4K & 0.578 & 0.609 & \textbf{0.704} & 0.653 \\
                w. CoT  & 4K & \textbf{0.712} & \textbf{0.754} & \textbf{0.704} & \textbf{0.728} \\
                \bottomrule
            \end{tabular}
        }
        \caption{CoT substantially improves SWR reward prediction quality.}
        \label{tab:cot_swr}
    \end{minipage}
\end{table}

We observe a distinct asymmetry in the benefits of Chain-of-Thought (CoT) reasoning between the transition and reward models:

\paratitle{Marginal Gain for SWT.}
As shown in Table~\ref{tab:cot_swt}, adding CoT to the transition model yields only a negligible improvement (0.8\%) in the downstream agent's resolve rate. We attribute this to the informational robustness of the transition task. The primary role of SWT is to simulate textual feedback (\ie \texttt{stdout}, \texttt{stderr}, \texttt{exit\_code}). Small inconsistencies in such feedback are often tolerable: the agent can still interpret the signal, continue exploring, and correct itself via subsequent actions and verification. Since CoT can substantially increase inference latency, the modest fidelity gain is typically not worth the added cost for step-level simulation.

\paratitle{Significant Gain for SWR.}
In contrast, Table~\ref{tab:cot_swr} shows that CoT is critical for the reward model, boosting accuracy by over 13\% (0.578 $\to$ 0.712) and Precision significantly. This is due to the binary sensitivity of the reward task. The SWR must synthesize complex test execution report into a strictly binary signal (Success/Failure). A minor misinterpretation of a test report can flip the reward label. CoT encourages more complete reasoning over repository context, patch effects, and test logic, thereby improving the fidelity of the predicted test report and making the resulting reward assignment substantially more accurate. We further analyze how this CoT integration impacts the stability of RL training in Section~\ref{subsec:rl_dynamics}.

\subsection{RL Dynamics}
\label{subsec:rl_dynamics}

\begin{figure}[t]
    \centering
    \includegraphics[width=1\linewidth]{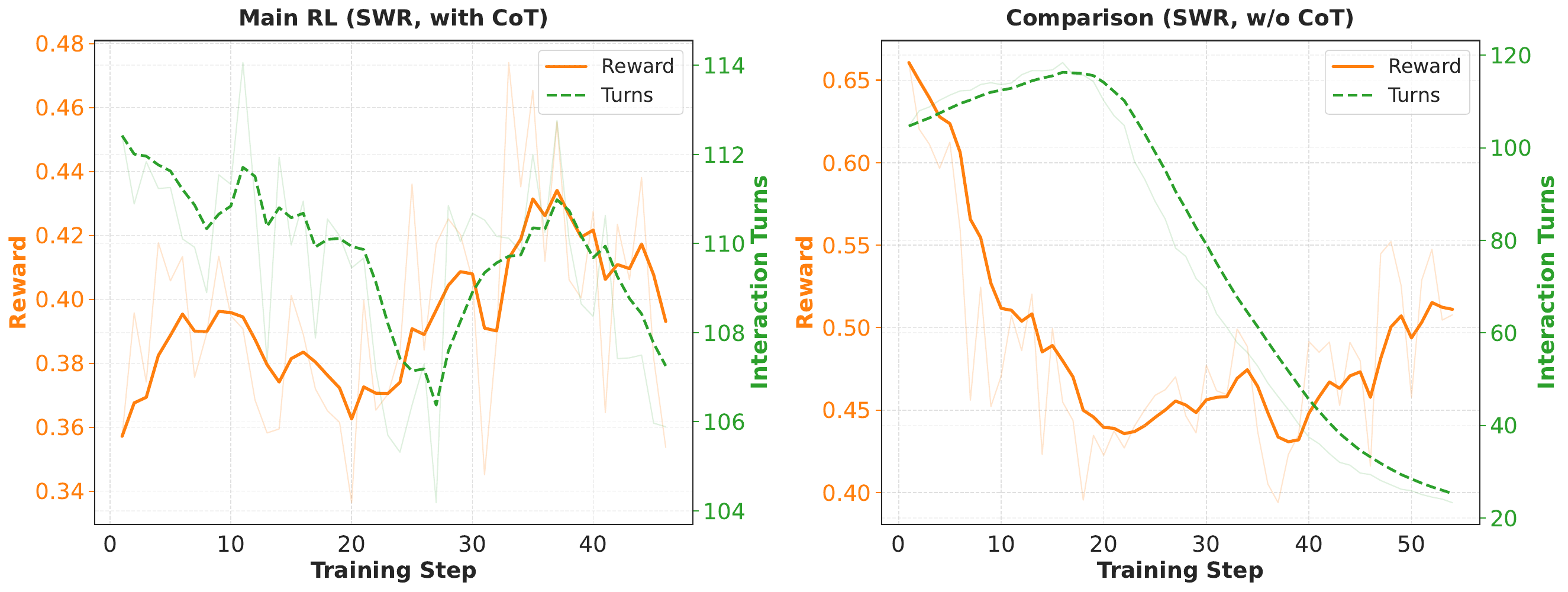}
    \caption{RL training dynamics using SWT-32B. Orange lines denote average reward; Green dashed lines denote mean interaction turns. Left: Main experiment using CoT-enhanced SWR-32B shows stable learning. Right: Comparison with non-CoT SWR-32B leads to trajectory length collapse, indicating reward hacking.}
    \label{fig:reward_vs_steps_main}
\end{figure}

\begin{figure}[t]
    \centering
    \includegraphics[width=0.7\linewidth]{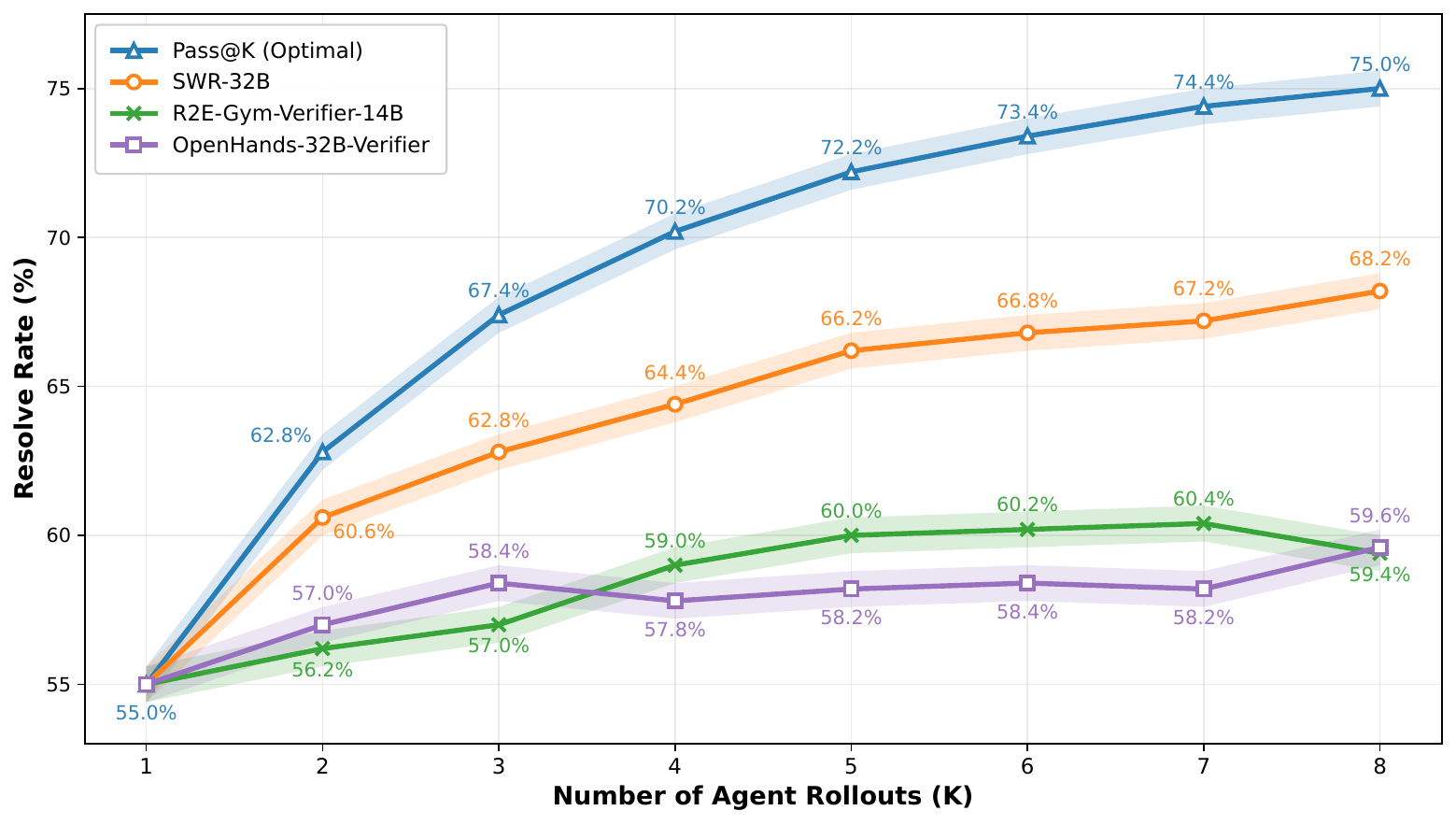}
    \caption{Test-time scaling on SWE-bench Verified: comparing SWR-32B with prior verifiers.}
    \label{fig:tts}
\end{figure}

We analyze the stability of our reinforcement learning process by tracking the reward and interaction turns over training steps. Figure~\ref{fig:reward_vs_steps_main} contrasts our main RL training with a comparison run under an alternative reward model setting.

\paratitle{Main RL Training.}
In the main run (Left), the agent exhibits healthy optimization dynamics: the reward steadily increases while the interaction length shows a gentle, efficiency-driven decline. This suggests that the policy performance improves steadily over training, without collapsing into degenerate behaviors.

\paratitle{Comparison Run.}
In contrast, the comparison run (Right), which uses a non-CoT SWR, exhibits clear signs of \emph{reward hacking}. The trajectory length collapses drastically after step 20, as the policy learns to exploit the reward model's low precision by submitting short, invalid solutions that are mistakenly flagged as correct.
These results suggest that a robust and faithful reward signal is critical for stable Docker-free reinforcement learning, and CoT provides an effective way to enhance it.

\subsection{Test-Time Scaling}
\label{sec:ana_tts}

Figure~\ref{fig:tts} compares test-time scaling on SWE-bench Verified using agent rollouts generated by SWT-32B, where we rank $K$ candidate trajectories with a verifier and select the best submission. 

\paratitle{Superior Scaling Performance.}
As shown in Figure~\ref{fig:tts}, SWR-32B delivers strong gains under TTS: performance increases from 55.0 at $K{=}1$ to 68.2 at TTS@8, an absolute improvement of 13.2\%, and substantially outperforms both R2E-Gym-Verifier-14B (59.4 at TTS@8) and OpenHands-32B-Verifier (59.6 at TTS@8).
Notably, the gap between SWR-32B and the Pass@$K$ (Optimal) upper bound is much smaller than that of prior verifiers, indicating stronger reward modeling and a more effective ranking signal.
Moreover, SWR-32B improves monotonically from $K=1$ to $K=8$, indicating a promising scaling behavior.

\paratitle{Generative Verification vs. Regression.}
Beyond the final score, the scaling trend reveals a fundamental difference in signal fidelity.
SWR-32B improves steadily from $K=1$ to $K=8$, whereas R2E-Gym-Verifier-14B plateaus early (around TTS@5) and yields little benefit thereafter; OpenHands-32B-Verifier is even less stable, showing noticeable fluctuations after $K{\ge}3$.
This contrast highlights the advantage of our generative simulation paradigm over token-level scoring.
By functioning as a virtual test runner that explicitly predicts a structured test report based on the evaluation context, SWR grounds its judgment in fine-grained execution logic, enabling it to robustly distinguish truly correct fixes from near-misses even within a large candidate pool.
In contrast, baselines that bypass this reasoning to derive scores directly from trajectory-level \textsc{Yes}/\textsc{No} log-probabilities (via softmax) are highly susceptible to the noise inherent in long interaction histories.
Lacking explicit verification steps, these ``black-box'' signals lose discrimination power as the candidate set grows, leading to early saturation and unstable scaling.

\subsection{Qualitative Fidelity of SWT and SWR Simulation}
\label{subsec:qual_fidelity}

\begin{figure}[t]
    \centering
    \includegraphics[width=1.0\linewidth]{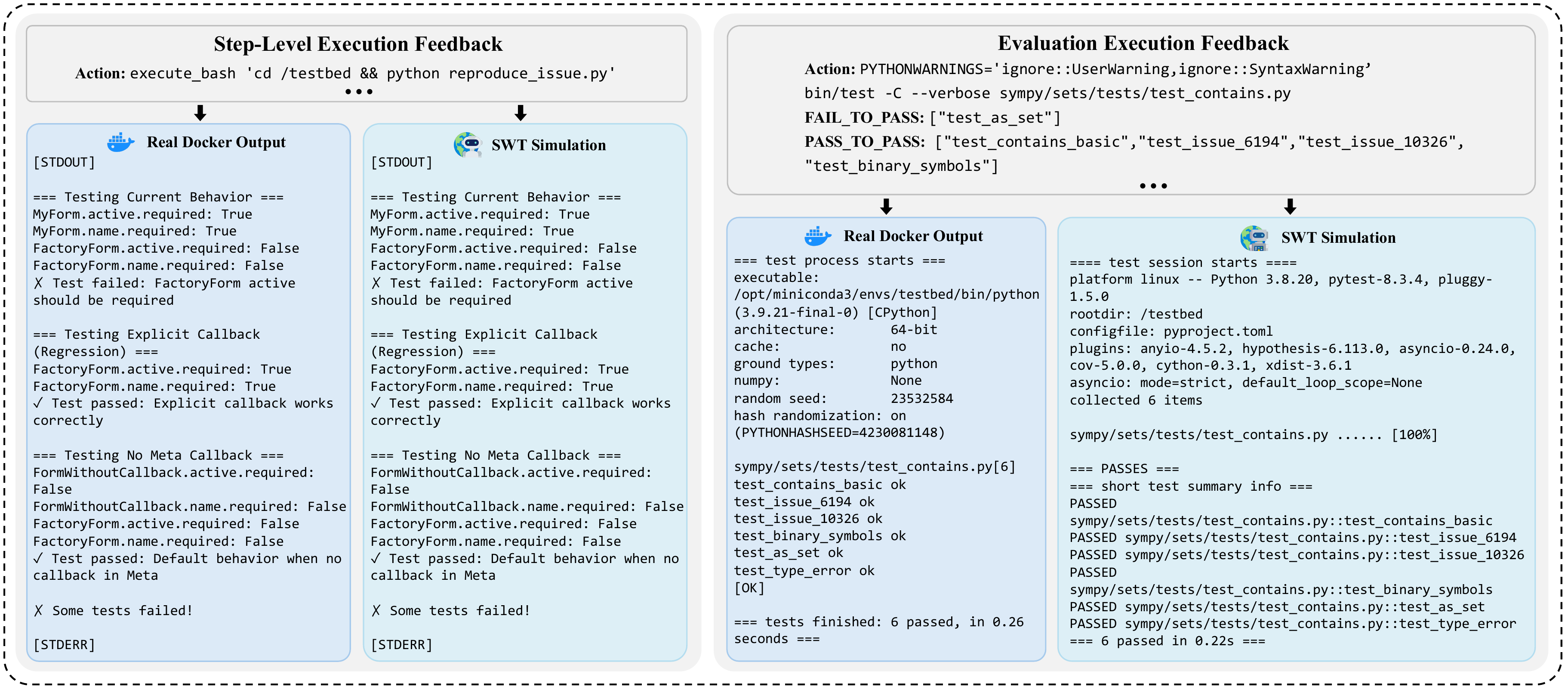}
    \caption{Qualitative fidelity of SWE-World simulation.}
    \label{fig:case}
\end{figure}

We qualitatively validate the fidelity of SWE-World feedback by comparing Docker ground-truth outputs with SWT/SWR predictions under the same context (Figure~\ref{fig:case}).
For readability, we omit lengthy fields in the input context and only present the essential information: (i) the command~(action) to be simulated for SWT, and (ii) the test command~(action) together with the F2P/P2P test sets for SWR.

\paratitle{SWT: Faithful Step-Level Execution Feedback.}
In the SWT example, the agent runs \texttt{python reproduce\_issue.py} to reproduce a Django issue.
The simulated \texttt{stdout} matches the real output almost \emph{line-by-line}, capturing the printed field requirements, the failure mode, and the assertion error message and exit status.
It also preserves the script's structure, indicating that SWT emulates the causal consequence of the current code state under the given command rather than producing generic-looking feedback.

\paratitle{SWR: Semantically Consistent Test-Based Verification.}
For SWR, we compare real test execution with SWR's predicted report on a SymPy instance.

There are some formatting differences between the real test report and the output generated by SWR, as the training data adopts a unified template that produces standardized \texttt{pytest}-style reports. Despite the format shift, the semantics align: SWR correctly predicts that all collected tests pass and explicitly lists key ones, including the F2P target and the P2P regression set.
Non-essential artifacts (\eg version strings and wall-clock time) may differ, but the core verification signal is preserved.

Overall, these examples suggest that SWE-World produces realistic and trustworthy feedback: SWT closely matches real step outputs, and SWR provides content-faithful test-based verification under a standardized report format.

\section{Conclusion}

In this paper, we propose SWE-World, a Docker-free framework (\ie execution-free) for training code agents. By training and leveraging repository-level environment and reward simulation models (\ie SWT and SWR), we establish a fully Docker-free pipeline for data synthesis, supervised fine-tuning, and reinforcement learning. This approach effectively circumvents the deployment complexity and concurrency bottlenecks inherent in traditional Docker-based trajectory generation. Evaluation results on the SWE-bench-Verified dataset demonstrate that agents trained within this framework achieve performance comparable to those trained with ground-truth execution feedback. Furthermore, the SWR model proves highly effective for test-time scaling, providing a grounded verification mechanism to select superior solutions without the need to execute actual unit tests. Additionally, by eliminating the strict dependency on buildable environments, SWE-World unlocks vast amounts of previously inaccessible open-source data resources, such as unbuildable Pull Requests and issues.

\bibliographystyle{unsrt}
\bibliography{ref.bib}

@article{wang2025openhands,
  title={The OpenHands Software Agent SDK: A Composable and Extensible Foundation for Production Agents},
  author={Wang, Xingyao and Rosenberg, Simon and Michelini, Juan and Smith, Calvin and Tran, Hoang and Nyst, Engel and Malhotra, Rohit and Zhou, Xuhui and Chen, Valerie and Brennan, Robert and others},
  journal={arXiv preprint arXiv:2511.03690},
  year={2025}
}

@article{jain2024livecodebench,
  title={Livecodebench: Holistic and contamination free evaluation of large language models for code},
  author={Jain, Naman and Han, King and Gu, Alex and Li, Wen-Ding and Yan, Fanjia and Zhang, Tianjun and Wang, Sida and Solar-Lezama, Armando and Sen, Koushik and Stoica, Ion},
  journal={arXiv preprint arXiv:2403.07974},
  year={2024}
}

@article{qwen2025qwen25technicalreport,
  author       = {An Yang and
                  Baosong Yang and
                  Beichen Zhang and
                  Binyuan Hui and
                  Bo Zheng and
                  Bowen Yu and
                  Chengyuan Li and
                  Dayiheng Liu and
                  Fei Huang and
                  Haoran Wei and
                  Huan Lin and
                  Jian Yang and
                  Jianhong Tu and
                  Jianwei Zhang and
                  Jianxin Yang and
                  Jiaxi Yang and
                  Jingren Zhou and
                  Junyang Lin and
                  Kai Dang and
                  Keming Lu and
                  Keqin Bao and
                  Kexin Yang and
                  Le Yu and
                  Mei Li and
                  Mingfeng Xue and
                  Pei Zhang and
                  Qin Zhu and
                  Rui Men and
                  Runji Lin and
                  Tianhao Li and
                  Tingyu Xia and
                  Xingzhang Ren and
                  Xuancheng Ren and
                  Yang Fan and
                  Yang Su and
                  Yichang Zhang and
                  Yu Wan and
                  Yuqiong Liu and
                  Zeyu Cui and
                  Zhenru Zhang and
                  Zihan Qiu},
  title        = {Qwen2.5 Technical Report},
  journal      = {CoRR},
  volume       = {abs/2412.15115},
  year         = {2024},
  url          = {https://doi.org/10.48550/arXiv.2412.15115},
  doi          = {10.48550/ARXIV.2412.15115},
  eprinttype    = {arXiv},
  eprint       = {2412.15115},
  bibsource    = {dblp computer science bibliography, https://dblp.org}
}

@article{chen2021evaluating,
  title={Evaluating large language models trained on code},
  author={Chen, Mark},
  journal={arXiv preprint arXiv:2107.03374},
  year={2021}
}

@misc{yang2025qwen3technicalreport,
      title={Qwen3 Technical Report}, 
      author={An Yang and Anfeng Li and Baosong Yang and Beichen Zhang and Binyuan Hui and Bo Zheng and Bowen Yu and Chang Gao and Chengen Huang and Chenxu Lv and Chujie Zheng and Dayiheng Liu and Fan Zhou and Fei Huang and Feng Hu and Hao Ge and Haoran Wei and Huan Lin and Jialong Tang and Jian Yang and Jianhong Tu and Jianwei Zhang and Jianxin Yang and Jiaxi Yang and Jing Zhou and Jingren Zhou and Junyang Lin and Kai Dang and Keqin Bao and Kexin Yang and Le Yu and Lianghao Deng and Mei Li and Mingfeng Xue and Mingze Li and Pei Zhang and Peng Wang and Qin Zhu and Rui Men and Ruize Gao and Shixuan Liu and Shuang Luo and Tianhao Li and Tianyi Tang and Wenbiao Yin and Xingzhang Ren and Xinyu Wang and Xinyu Zhang and Xuancheng Ren and Yang Fan and Yang Su and Yichang Zhang and Yinger Zhang and Yu Wan and Yuqiong Liu and Zekun Wang and Zeyu Cui and Zhenru Zhang and Zhipeng Zhou and Zihan Qiu},
      year={2025},
      eprint={2505.09388},
      archivePrefix={arXiv},
      primaryClass={cs.CL},
      url={https://arxiv.org/abs/2505.09388}, 
}

@article{huang2025transformers,
  title={Transformers provably learn chain-of-thought reasoning with length generalization},
  author={Huang, Yu and Wen, Zixin and Singh, Aarti and Chi, Yuejie and Chen, Yuxin},
  journal={arXiv preprint arXiv:2511.07378},
  year={2025}
}

@misc{sweb-verified,
    title = {Introducing SWE-bench Verified},
    url = {https://openai.com/index/introducing-swe-bench-verified/},
    author = {Neil Chowdhury and James Aung and Chan Jun Shern and Oliver Jaffe and Dane Sherburn and Giulio Starace and Evan Mays and Rachel Dias and Marwan Aljubeh and Mia Glaese and Carlos E. Jimenez and John Yang and Leyton Ho and Tejal Patwardhan and Kevin Liu and Aleksander Madry},
    year={2024},
    month = {August},
}

@article{xiao2026mimo,
  title={MiMo-V2-Flash Technical Report},
  author={Xiao, Bangjun and Xia, Bingquan and Yang, Bo and Gao, Bofei and Shen, Bowen and Zhang, Chen and He, Chenhong and Lou, Chiheng and Luo, Fuli and Wang, Gang and others},
  journal={arXiv preprint arXiv:2601.02780},
  year={2026}
}

@misc{sun2025simpledeepsearcherdeepinformationseeking,
      title={SimpleDeepSearcher: Deep Information Seeking via Web-Powered Reasoning Trajectory Synthesis}, 
      author={Shuang Sun and Huatong Song and Yuhao Wang and Ruiyang Ren and Jinhao Jiang and Junjie Zhang and Fei Bai and Jia Deng and Wayne Xin Zhao and Zheng Liu and Lei Fang and Zhongyuan Wang and Ji-Rong Wen},
      year={2025},
      eprint={2505.16834},
      archivePrefix={arXiv},
      primaryClass={cs.CL},
      url={https://arxiv.org/abs/2505.16834}, 
}

@article{xie2025swe,
  title={Swe-fixer: Training open-source llms for effective and efficient github issue resolution},
  author={Xie, Chengxing and Li, Bowen and Gao, Chang and Du, He and Lam, Wai and Zou, Difan and Chen, Kai},
  journal={arXiv preprint arXiv:2501.05040},
  year={2025}
}

@misc{SWESwiss2025,
    title={SWE-Swiss: A Multi-Task Fine-Tuning and RL Recipe for High-Performance Issue Resolution},
    author={He, Zhenyu and Yang, Qingping and Sheng, Wei and Zhong, Xiaojian and Zhang, Kechi and An, Chenxin and Shi, Wenlei and Cai, Tianle and He, Di and Chen, Jiaze and Xu, Jingjing},
    howpublished={\url{https://github.com/zhenyuhe00/SWE-Swiss}},
    note={Notion Blog},
    year={2025}
}

@article{yang2024swe,
  title={Swe-agent: Agent-computer interfaces enable automated software engineering},
  author={Yang, John and Jimenez, Carlos E and Wettig, Alexander and Lieret, Kilian and Yao, Shunyu and Narasimhan, Karthik and Press, Ofir},
  journal={Advances in Neural Information Processing Systems},
  volume={37},
  pages={50528--50652},
  year={2024}
}

@misc{minimax_m2,
  title        = {MiniMax M2 \& Agent: Ingenious in Simplicity},
  author       = {{MiniMax}},
  howpublished = {\url{https://www.minimax.io/news/minimax-m2}},
  note         = {2025-10-27},
  year         = {2025}
}

@misc{minimax_m21,
  title        = {M2.1: Multilingual and Multi-Task Coding with Strong Generalization},
  author       = {{MiniMax}},
  howpublished = {\url{https://www.minimaxi.com/news/m21-multilingual-and-multi-task-coding-with-strong-general}},
  note         = {2025-01-04},
  year         = {2026}
}

@misc{glm_46,
  title        = {GLM-4.6: Advanced Agentic, Reasoning and Coding Capabilities},
  author       = {{Z.ai}},
  howpublished = {\url{https://z.ai/blog/glm-4.6}},
  note         = {2025-09-30},
  year         = {2025}
}

@article{jain2025r2e,
  title={R2e-gym: Procedural environments and hybrid verifiers for scaling open-weights swe agents},
  author={Jain, Naman and Singh, Jaskirat and Shetty, Manish and Zheng, Liang and Sen, Koushik and Stoica, Ion},
  journal={arXiv preprint arXiv:2504.07164},
  year={2025}
}

@article{hui2024qwen2,
  title={Qwen2. 5-coder technical report},
  author={Hui, Binyuan and Yang, Jian and Cui, Zeyu and Yang, Jiaxi and Liu, Dayiheng and Zhang, Lei and Liu, Tianyu and Zhang, Jiajun and Yu, Bowen and Lu, Keming and others},
  journal={arXiv preprint arXiv:2409.12186},
  year={2024}
}

@inproceedings{yao2022react,
  title={React: Synergizing reasoning and acting in language models},
  author={Yao, Shunyu and Zhao, Jeffrey and Yu, Dian and Du, Nan and Shafran, Izhak and Narasimhan, Karthik R and Cao, Yuan},
  booktitle={The eleventh international conference on learning representations},
  year={2022}
}

@inproceedings{jimenez2024swebench,
  title={SWE-bench: Can Language Models Resolve Real-World GitHub Issues?},
  author={Jimenez, Carlos E and Yang, John and Wettig, Alexander and Yao, Shunyu and Pei, Kexin and Press, Ofir and Narasimhan, Karthik},
  booktitle={The Twelfth International Conference on Learning Representations},
  year={2024}
}

@misc{yang2025swesmith,
      title={SWE-smith: Scaling Data for Software Engineering Agents}, 
      author={John Yang and Kilian Lieret and Carlos E. Jimenez and Alexander Wettig and Kabir Khandpur and Yanzhe Zhang and Binyuan Hui and Ofir Press and Ludwig Schmidt and Diyi Yang},
      year={2025},
      eprint={2504.21798},
      archivePrefix={arXiv},
      primaryClass={cs.SE}
}

@misc{badertdinov2025swerebench,
      title={SWE-rebench: An Automated Pipeline for Task Collection and Decontaminated Evaluation of Software Engineering Agents}, 
      author={Ibragim Badertdinov and Alexander Golubev and Maksim Nekrashevich and Anton Shevtsov and Simon Karasik and Andrei Andriushchenko and Maria Trofimova and Daria Litvintseva and Boris Yangel},
      year={2025},
      eprint={2505.20411},
      archivePrefix={arXiv},
      primaryClass={cs.SE}
}

@misc{xia2024agentless,
      title={Agentless: Demystifying LLM-based Software Engineering Agents}, 
      author={Chunqiu Steven Xia and Yinlin Deng and Soren Dunn and Lingming Zhang},
      year={2024},
      eprint={2407.01489},
      archivePrefix={arXiv},
      primaryClass={cs.SE}
}

@misc{minisweagent,
  author = {SWE-agent Team},
  title = {Mini-SWE-Agent},
  year = {2024},
  publisher = {GitHub},
  journal = {GitHub repository},
  howpublished = {\url{https://github.com/SWE-agent/Mini-SWE-Agent}}
}

@article{sonwane2025bugpilot,
  title={BugPilot: Complex Bug Generation for Efficient Learning of SWE Skills},
  author={Sonwane, Atharv and White, Isadora and Lee, Hyunji and Pereira, Matheus and Caccia, Lucas and Kim, Minseon and Shi, Zhengyan and Singh, Chinmay and Sordoni, Alessandro and C{\^o}t{\'e}, Marc-Alexandre and others},
  journal={arXiv preprint arXiv:2510.19898},
  year={2025}
}

@article{zeng2025skywork,
  title={Skywork-SWE: Unveiling Data Scaling Laws for Software Engineering in LLMs},
  author={Zeng, Liang and Li, Yongcong and Xiao, Yuzhen and Li, Changshi and Liu, Chris Yuhao and Yan, Rui and Wei, Tianwen and He, Jujie and Song, Xuchen and Liu, Yang and others},
  journal={arXiv preprint arXiv:2506.19290},
  year={2025}
}

@inproceedings{swe-gym,
  author       = {Jiayi Pan and
                  Xingyao Wang and
                  Graham Neubig and
                  Navdeep Jaitly and
                  Heng Ji and
                  Alane Suhr and
                  Yizhe Zhang},
  title        = {Training Software Engineering Agents and Verifiers with SWE-Gym},
  booktitle    = {Forty-second International Conference on Machine Learning, {ICML}
                  2025, Vancouver, BC, Canada, July 13-19, 2025},
  publisher    = {OpenReview.net},
  year         = {2025},
  url          = {https://openreview.net/forum?id=Cq1BNvHx74},
  bibsource    = {dblp computer science bibliography, https://dblp.org}
}

@misc{deepswe2025,
  title={DeepSWE: Training a Fully Open-sourced, State-of-the-Art Coding Agent by Scaling RL},
  author={Luo, Michael and Jain, Naman and Singh, Jaskirat and Tan, Sijun and Patel, Ameen and Wu, Qingyang and Ariyak, Alpay and Cai, Colin and Venkat, Tarun and Zhu, Shang and Athiwaratkun, Ben and Roongta, Manan and Zhang, Ce and Li, Li Erran and Popa, Raluca Ada and Sen, Koushik and Stoica, Ion},
  howpublished={\url{https://www.together.ai/blog/deepswe}},
  note={Blog post},
  year={2025},
  month={7},
  url={https://www.together.ai/blog/deepswe}
}

@article{copet2025cwm,
  title={CWM: An Open-Weights LLM for Research on Code Generation with World Models},
  author={Copet, Jade and Carbonneaux, Quentin and Cohen, Gal and Gehring, Jonas and Kahn, Jacob and Kossen, Jannik and Kreuk, Felix and McMilin, Emily and Meyer, Michel and Wei, Yuxiang and others},
  journal={arXiv preprint arXiv:2510.02387},
  year={2025}
}

@article{golubev2025training,
  title={Training long-context, multi-turn software engineering agents with reinforcement learning},
  author={Golubev, Alexander and Trofimova, Maria and Polezhaev, Sergei and Badertdinov, Ibragim and Nekrashevich, Maksim and Shevtsov, Anton and Karasik, Simon and Abramov, Sergey and Andriushchenko, Andrei and Fisin, Filipp and others},
  journal={arXiv preprint arXiv:2508.03501},
  year={2025}
}

@article{yang2025kimi,
  title={Kimi-Dev: Agentless Training as Skill Prior for SWE-Agents},
  author={Yang, Zonghan and Wang, Shengjie and Fu, Kelin and He, Wenyang and Xiong, Weimin and Liu, Yibo and Miao, Yibo and Gao, Bofei and Wang, Yejie and Ma, Yingwei and others},
  journal={arXiv preprint arXiv:2509.23045},
  year={2025}
}

@article{tao2026swe,
  title={SWE-Lego: Pushing the Limits of Supervised Fine-tuning for Software Issue Resolving},
  author={Tao, Chaofan and Chen, Jierun and Jiang, Yuxin and Kou, Kaiqi and Wang, Shaowei and Wang, Ruoyu and Li, Xiaohui and Yang, Sidi and Du, Yiming and Dai, Jianbo and others},
  journal={arXiv preprint arXiv:2601.01426},
  year={2026}
}

@article{shao2024deepseekmath,
  title={Deepseekmath: Pushing the limits of mathematical reasoning in open language models},
  author={Shao, Zhihong and Wang, Peiyi and Zhu, Qihao and Xu, Runxin and Song, Junxiao and Bi, Xiao and Zhang, Haowei and Zhang, Mingchuan and Li, YK and Wu, Yang and others},
  journal={arXiv preprint arXiv:2402.03300},
  year={2024}
}

@article{liu2025context,
  title={Context as a Tool: Context Management for Long-Horizon SWE-Agents},
  author={Liu, Shukai and Yang, Jian and Jiang, Bo and Li, Yizhi and Guo, Jinyang and Liu, Xianglong and Dai, Bryan},
  journal={arXiv preprint arXiv:2512.22087},
  year={2025}
}

@article{wang2025swe,
  title={Swe-mirror: Scaling issue-resolving datasets by mirroring issues across repositories},
  author={Wang, Junhao and Zan, Daoguang and Xin, Shulin and Liu, Siyao and Wu, Yurong and Shen, Kai},
  journal={arXiv preprint arXiv:2509.08724},
  year={2025}
}

@article{zeng2026davinci,
  title={daVinci-Dev: Agent-native Mid-training for Software Engineering},
  author={Zeng, Ji and Fu, Dayuan and Mi, Tiantian and Zhuang, Yumin and Huang, Yaxing and Li, Xuefeng and Ye, Lyumanshan and Xie, Muhang and Hua, Qishuo and Huang, Zhen and others},
  journal={arXiv preprint arXiv:2601.18418},
  year={2026}
}

@article{shum2025swe,
  title={SWE-RM: Execution-free Feedback For Software Engineering Agents},
  author={Shum, KaShun and Hui, Binyuan and Chen, Jiawei and Zhang, Lei and Yang, Jiaxi and Huang, Yuzhen and Lin, Junyang and He, Junxian and others},
  journal={arXiv preprint arXiv:2512.21919},
  year={2025}
}

@article{cao2025skyrl,
  title={SkyRL-Agent: Efficient RL Training for Multi-turn LLM Agent},
  author={Cao, Shiyi and Li, Dacheng and Zhao, Fangzhou and Yuan, Shuo and Hegde, Sumanth R and Chen, Connor and Ruan, Charlie and Griggs, Tyler and Liu, Shu and Tang, Eric and others},
  journal={arXiv preprint arXiv:2511.16108},
  year={2025}
}

@article{song2025r1,
  title={R1-searcher: Incentivizing the search capability in llms via reinforcement learning},
  author={Song, Huatong and Jiang, Jinhao and Min, Yingqian and Chen, Jie and Chen, Zhipeng and Zhao, Wayne Xin and Fang, Lei and Wen, Ji-Rong},
  journal={arXiv preprint arXiv:2503.05592},
  year={2025}
}

@article{zhao2023survey,
  title={A survey of large language models},
  author={Zhao, Wayne Xin and Zhou, Kun and Li, Junyi and Tang, Tianyi and Wang, Xiaolei and Hou, Yupeng and Min, Yingqian and Zhang, Beichen and Zhang, Junjie and Dong, Zican and others},
  journal={arXiv preprint arXiv:2303.18223},
  volume={1},
  number={2},
  year={2023}
}

@article{luo2025mcp,
  title={Mcp-universe: Benchmarking large language models with real-world model context protocol servers},
  author={Luo, Ziyang and Shen, Zhiqi and Yang, Wenzhuo and Zhao, Zirui and Jwalapuram, Prathyusha and Saha, Amrita and Sahoo, Doyen and Savarese, Silvio and Xiong, Caiming and Li, Junnan},
  journal={arXiv preprint arXiv:2508.14704},
  year={2025}
}

@article{kojima2022large,
  title={Large language models are zero-shot reasoners},
  author={Kojima, Takeshi and Gu, Shixiang Shane and Reid, Machel and Matsuo, Yutaka and Iwasawa, Yusuke},
  journal={Advances in neural information processing systems},
  volume={35},
  pages={22199--22213},
  year={2022}
}

@article{wei2022chain,
  title={Chain-of-thought prompting elicits reasoning in large language models},
  author={Wei, Jason and Wang, Xuezhi and Schuurmans, Dale and Bosma, Maarten and Xia, Fei and Chi, Ed and Le, Quoc V and Zhou, Denny and others},
  journal={Advances in neural information processing systems},
  volume={35},
  pages={24824--24837},
  year={2022}
}

@article{gu2024survey,
  title={A survey on llm-as-a-judge},
  author={Gu, Jiawei and Jiang, Xuhui and Shi, Zhichao and Tan, Hexiang and Zhai, Xuehao and Xu, Chengjin and Li, Wei and Shen, Yinghan and Ma, Shengjie and Liu, Honghao and others},
  journal={The Innovation},
  year={2024},
  publisher={Elsevier}
}

@inproceedings{chen2025reverse,
  title={Reverse thinking makes llms stronger reasoners},
  author={Chen, Justin and Wang, Zifeng and Palangi, Hamid and Han, Rujun and Ebrahimi, Sayna and Le, Long and Perot, Vincent and Mishra, Swaroop and Bansal, Mohit and Lee, Chen-Yu and others},
  booktitle={Proceedings of the 2025 Conference of the Nations of the Americas Chapter of the Association for Computational Linguistics: Human Language Technologies (Volume 1: Long Papers)},
  pages={8611--8630},
  year={2025}
}

@article{wei2025swe,
  title={Swe-rl: Advancing llm reasoning via reinforcement learning on open software evolution},
  author={Wei, Yuxiang and Duchenne, Olivier and Copet, Jade and Carbonneaux, Quentin and Zhang, Lingming and Fried, Daniel and Synnaeve, Gabriel and Singh, Rishabh and Wang, Sida I},
  journal={arXiv preprint arXiv:2502.18449},
  year={2025}
}

@article{ma2024lingma,
  title={Lingma swe-gpt: An open development-process-centric language model for automated software improvement},
  author={Ma, Yingwei and Cao, Rongyu and Cao, Yongchang and Zhang, Yue and Chen, Jue and Liu, Yibo and Liu, Yuchen and Li, Binhua and Huang, Fei and Li, Yongbin},
  journal={arXiv preprint arXiv:2411.00622},
  year={2024}
}

@article{lin2025scaling,
  title={Scaling Code-Assisted Chain-of-Thoughts and Instructions for Model Reasoning},
  author={Lin, Honglin and Pei, Qizhi and Gao, Xin and Pan, Zhuoshi and Li, Yu and Li, Juntao and He, Conghui and Wu, Lijun},
  journal={arXiv preprint arXiv:2510.04081},
  year={2025}
}

\newpage
\appendix
\section{Instance Metadata and Context Fields}
\label{app:instance_format}

This section details the instance-level metadata used throughout the paper, including the task definition $\mathcal{I}=(R,b,d,\mathcal{U})$ and the fields involved in constructing SWT/SWR contexts.

\subsection{Instance Metadata Schema}
\label{app:instance_metadata_schema}

\begin{table}[ht]
    \small
    \centering
    \setlength{\tabcolsep}{6pt}
    \renewcommand{\arraystretch}{1.15}
    \resizebox{\linewidth}{!}{%
    \begin{tabular}{p{0.22\linewidth} p{0.74\linewidth}}
        \toprule
        \textbf{Key} & \textbf{Description} \\
        \midrule
        \texttt{repo} &
        Repository name from which the task is sourced. \\

        \texttt{instance\_id} &
        A unique identifier constructed from the repository and its PR/issue ID. \\

        \texttt{base\_commit} &
        Commit hash of the reference repository snapshot used to instantiate the task. \\

        \texttt{hints\_text} &
        Optional natural-language hints that guide the intended fix. \\

        \texttt{problem\_statement} &
        Natural-language specification of the desired change (bug report / feature request). \\

        \texttt{FAIL\_TO\_PASS} &
        Unit tests expected to flip from failing to passing after a correct fix (F2P). \\

        \texttt{PASS\_TO\_PASS} &
        Regression unit tests that must remain passing after the fix (P2P). \\

        \texttt{gold\_patch} &
        Reference solution in \texttt{.patch} format, used internally for supervision and analysis. \\

        \texttt{test\_patch} &
        \texttt{.patch}-format string that adds hidden/unseen evaluation tests. \\
        \bottomrule
    \end{tabular}%
    }
    \caption{Instance metadata fields used to define $\mathcal{I}=(R,b,d,\mathcal{U})$ and to construct SWT/SWR contexts.}
    \label{tab:instance_metadata}
\end{table}

Table~\ref{tab:instance_metadata} summarizes the key metadata fields defining each SWE task instance and used to construct the SWT and SWR contexts.

\subsection{Instance Meta Data in SWT/SWR Contexts}
\label{app:context_fields}

\paratitle{Initial Analysis Generation.}
The \emph{Initial Analysis} used in both SWT and SWR contexts is generated from the instance metadata using GLM-4.6.
This analysis is designed to help SWT and SWR quickly grasp the core issue and the expected fix direction.
The prompt template is shown in Figure~\ref{p:init_analysis_prompt}.

\paratitle{SWT Context Meta Data.}
At step $t$, SWT takes a context $\kappa^{\text{SWT}}_t$.
Its \emph{Instance Meta Data} includes: (i) the problem statement \texttt{problem\_statement} ($d$), (ii) an \emph{Initial Analysis} summarizing the failure behavior, likely root cause, and intended fix, and (iii) the reference solution \texttt{gold\_patch} as an internal patch $P^{\star}$ for comparison, which is never exposed to the code agent.
In addition, $\kappa^{\text{SWT}}_t$ contains the agent's current patch $P_t$ and the execution content needed to simulate the command $a_t$. The prompt template for SWT is shown in Figure~\ref{p:swt_prompt}.

\paratitle{SWR Context Meta Data.}
SWR operates on an evaluation context $\kappa^{\text{SWR}}_{\text{eval}}$, which follows the same three-part structure as $\kappa^{\text{SWT}}_t$ (Instance Meta Data, Agent Patch, and Execution Content) and additionally includes the unit tests $\mathcal{U}$, consisting of \texttt{FAIL\_TO\_PASS} and \texttt{PASS\_TO\_PASS}.
For $\kappa^{\text{SWR}}_{\text{eval}}$, the Agent Patch corresponds to the final submitted patch $P$, and the Execution Content corresponds to the unit-test command together with the test content in $\mathcal{U}$. The prompt template for SWR is shown in Figure~\ref{p:swr_prompt}

\section{Training Details for SWT and SWR}
\label{app:training_details}

\subsection{Detailed Training Setup}
\label{app:detailed_training_setup}

\paratitle{Data Collection.}
We collect training data using the R2E-Gym framework, with at most 100 interaction turns per rollout.
Our task pool is built upon three open-source SWE datasets: SWE-Gym~\cite{swe-gym}, SWE-rebench~\cite{badertdinov2025swerebench}, and R2E-Gym.
We use powerful LLMs (GLM-4.6 and MinMax-M2; temperature 0.7) to interact with Docker-based environments and record real step-level execution outputs, final unit test reports and rewards.

Importantly, we use \emph{both successful and failed} rollouts: regardless of whether the final reward is $0$ or $1$, a trajectory is useful as long as we can extract the corresponding context and the \emph{ground-truth} Docker outputs for supervision.
We additionally filter out trajectories whose Docker environments are corrupted (\eg broken images, dependency failures, or abnormal logs) that produce invalid or inconsistent outputs.
For SWR data, we enforce a balanced label distribution by subsampling to achieve a $1{:}1$ ratio between reward-$0$ and reward-$1$ examples.

\paratitle{Chain-of-Thought (CoT) Augmentation.}
To improve reasoning fidelity, we augment both SWT and SWR training outputs with chain-of-thought generated by Qwen3-235B-A22B-Thinking (prompt shown in Figure~\ref{p:cot_prompt}).
For each example, we produce two versions of supervision:
(i) a \emph{non-CoT} version with only the final structured output, and
(ii) a \emph{CoT} version that prepends a reasoning trace wrapped by \texttt{<think>} \ldots \texttt{</think>}.
After augmentation, we obtain training datasets:
SWT: 26K non-CoT + 26K CoT;
SWR: 21K non-CoT + 21K CoT.

\paratitle{Models and Baselines.}
We train SWT and SWR using Qwen2.5-32B-Instruct and Qwen2.5-72B-Instruct.
As prompt-engineering baselines for both SWT and SWR, we adopt powerful open models (MinMax-M2.1 and GLM-4.7) to directly generate simulated feedback via carefully designed prompts.

\paratitle{Prompt Templates and Output Formats.}
We format each training example into a unified instruction-following template with explicit \texttt{input} fields.
The SWT and SWR input prompt templates are shown in Figure~\ref{p:swt_prompt} and Figure~\ref{p:swr_prompt}, respectively.
For outputs, we enforce a strict JSON format to facilitate reliable parsing:
\begin{itemize}[leftmargin=2.0em]
    \item SWT: \texttt{\{"stdout": ..., "stderr": ..., "exit\_code": ...\}}
    \item SWR: \texttt{\{"test\_report": ..., "reward": ...\}}
\end{itemize}
For CoT, we prepend the reasoning trace before the JSON:
\texttt{<think> ... </think> \{...\}}.
This design allows us to extract structured fields (\eg \texttt{stdout}/\texttt{exit\_code} or \texttt{reward}) without ambiguity.

\subsection{SFT Hyperparameters}
\label{app:sft_hparams}

Table~\ref{tab:sft_hparams} summarizes the main SFT hyperparameters extracted from our training commands.
All runs use AdamW and DeepSpeed ZeRO-3, with BF16, FlashAttention, ring attention, and gradient checkpointing enabled.
\begin{table}[t]
    \small
    \centering
    \setlength{\tabcolsep}{6pt}
    \resizebox{\linewidth}{!}{%
    \begin{tabular}{lcccccc}
        \toprule
        \textbf{Component} &
        \textbf{Global Batch Size} &
        \textbf{Max Len} &
        \textbf{LR} &
        \textbf{Scheduler} &
        \textbf{Warmup} &
        \textbf{Epochs} \\
        \midrule
        SWT-32B (non-CoT) &
        512 &
        98304 &
        $4{\times}10^{-5}$ &
        cosine &
        0.1 &
        4 \\
        SWT-32B (CoT) &
        512 &
        98304 &
        $4{\times}10^{-5}$ &
        cosine &
        0.1 &
        4 \\
        SWR-32B (CoT) &
        512 &
        98304 &
        $3{\times}10^{-5}$ &
        cosine &
        0.1 &
        2 \\
        \bottomrule
    \end{tabular}%
    }
    \caption{
        Key SFT hyperparameters for SWT/SWR training.
    }
    \label{tab:sft_hparams}
\end{table}

\paratitle{Final Trained Models.}
Considering resource constraints and training time, we train the following models used in the main paper:
SWT-32B (non-CoT; trained on 26K training examples),
SWT-32B-CoT (CoT; trained on 26K training examples),
SWT-72B-CoT (CoT; trained on 26K training examples),
SWR-32B (CoT; trained on 21K training examples),
SWR-72B (CoT; trained on 21K training examples).

\section{Training Details for Docker-Free SFT with SWE-World}

\label{app:ef_sft_details}

\paratitle{Data Collection with SWE-World Rollouts.}
We collect training data by rolling out code agents in a Docker-free setting using the R2E-Gym framework, with at most 100 interaction turns per rollout.
The rollout task pool includes three public SWE datasets---R2E-Gym, SWE-Gym, and SWE-rebench---together with our SWE-World dataset.
During rollouts, we replace the Docker backend with SWE-World: SWT provides step-level feedback, and SWR assigns the trajectory-level reward after termination.
For a favorable efficiency--accuracy trade-off, we use SWT-32B and SWR-32B with a 128K context window, using a sampling temperature of 0.
We drive rollouts with powerful LLM agents (GLM-4.6 and MinMax-M2; temperature 0.7) to generate diverse high-quality interaction traces.

\paratitle{Trajectory Filtering.}
We apply format-based filtering to ensure training stability.
We keep only successful trajectories (reward $=1$), and drop outliers that exceed 80K tokens or 100 turns to reduce OOM risks.
We further discard trajectories containing malformed actions (\eg unparsable tool/function calls or invalid repeated invocations).
After filtering, we obtain 5.7K SWE-World-only trajectories for Docker-free SFT.

\paratitle{Backbones.}
We fine-tune two policy backbones: Qwen2.5-32B-Coder-Instruct and Qwen3-4B-Instruct-2507.

\paratitle{SFT Hyperparameters.}
We train for 5 epochs with a maximum context length of 80K tokens and a global batch size of 256.
We use AdamW with a cosine learning-rate schedule, warming up for 10\% of training and decaying the learning rate from $5\times10^{-5}$ to $5\times10^{-6}$.

\section{Training Details for Docker-Free Agent RL with SWE-World}
\label{app:ef_rl_details}

\paratitle{Task Pool.}
We conduct reinforcement learning on a unified task pool constructed from three open-source SWE datasets: R2E-Gym, SWE-Gym, and SWE-rebench.

\paratitle{Execution-Free Rollouts with SWE-World.}
All RL rollouts are generated in the R2E-Gym framework with the Docker backend fully replaced by the SWE-World backend.
Concretely, we use SWT-32B to provide step-level execution feedback and SWR-32B to assign the terminal unit-test reward after the agent submits a patch.
Both SWT and SWR operate with a 128K context window, and we sample both models with a temperature of 0.

\paratitle{Policy Initialization.}
The RL policies are initialized from our Docker-free SFT checkpoints: SWE-World-4B-SFT and SWE-World-32B-SFT.

\paratitle{Optimization with GRPO++.}
We optimize the policy using GRPO++.
GRPO++ is a stabilized variant of group-based policy optimization that uses a clipped ratio objective with leave-one-out advantages and length normalization.
For completeness, we provide the objective and related definitions below.

\begingroup
\small
\begin{align}
\mathcal{J}(\theta)
&= \mathbb{E}_{ \{o_i\}_{i=1}^G \sim \pi_{\text{old}}(\cdot \mid q)}
\Bigg[
\frac{1}{G}\sum_{i=1}^G \frac{1}{L_{\max}} \sum_{t=1}^{|o_i|}
\min\Bigg(
\frac{\pi_\theta(o_{i,t}\mid q,o_{i,<t})}{\pi_{\text{old}}(o_{i,t}\mid q,o_{i,<t})}
\Bigg(R_i - \frac{1}{G-1}\!\!\sum_{\substack{j=1\\ j\neq i}}^G R_j\Bigg),
\nonumber \\
&\hspace{5.2em}
\text{clip}\!\left(
\frac{\pi_\theta(o_{i,t}\mid q,o_{i,<t})}{\pi_{\text{old}}(o_{i,t}\mid q,o_{i,<t})},
\,1-\varepsilon_{\text{low}},\,1+\varepsilon_{\text{high}}
\right)
\Bigg(R_i - \frac{1}{G-1}\!\!\sum_{\substack{j=1\\ j\neq i}}^G R_j\Bigg)
\Bigg)
\Bigg],
\end{align}
\endgroup
where $L_{\max}$ is a fixed constant.

\paratitle{Return Definition with SWR.}
Let $\hat{r}_i \in \{0,1\}$ denote the terminal reward predicted by SWR for rollout $o_i$.
We define the scalar return $R_i$ as
\begin{equation}
R_i=
\begin{cases}
\hat{r}_i, & \text{if the rollout ends with \texttt{submit}},\\
\alpha\,\hat{r}_i, & \text{otherwise},
\end{cases}
\qquad \alpha=0.5.
\end{equation}

\paratitle{RL Hyperparameters and Rollout Budget.}
We train with a constant learning rate of $1{\times}10^{-6}$ and optimize on batches of 32 problems per update.
For each problem, we sample 4 rollouts in parallel with a policy sampling temperature of 1.0.
To control cost and stabilize long-context training, each trajectory is capped at 150 interaction turns, uses a maximum context length of 108K tokens during training, and is subject to a per-trajectory timeout of 5{,}400 seconds.

\section{Additional Evaluation Configurations}
\label{app:eval_config}

This section provides the concrete inference configurations used in our evaluations.

\paratitle{SWT/SWR/TTS Inference Settings.}
For SWT Evaluation, SWR Evaluation, and Test-Time Scaling, we use a maximum context length of 128K tokens and set the sampling temperature to 0 for deterministic model outputs.

\paratitle{SWE Agent Evaluation Settings.}
For SWE agent evaluation, we run the agent with a sampling temperature of 0.7, a maximum context length of 128K tokens, and a maximum of 150 interaction turns.
All final correctness judgments are obtained by running the submitted patch in the Docker evaluation harness.

\section{SWE-World Dataset Construction and Statistics}
\label{app:dataset_stats}

We construct SWE-World dataset by extracting Python Issue-PR pairs from the GitHub Archive covering the period from 2010 to 2025. To ensure the semantic richness and quality required for downstream tasks, we apply a rigorous filtering pipeline.

\subsection{Data Selection Heuristics}
\label{app:data_heuristics}

We first utilize regex matching to eliminate bot-generated noise. We retain issues that contain bug-related keywords and meet specific length constraints: titles must exceed 20 characters and issue bodies must exceed 200 characters. Additionally, we require at least three high-quality community comments to ensure sufficient context.

For the linked Pull Requests (PRs), we apply the following structural constraints to ensure the tasks are solvable yet non-trivial:
\begin{itemize}[leftmargin=2.0em]
    \item File Count: The number of modified files ranges from 1 to 20.
    \item Code Churn: The total code churn (additions + deletions) is between 1 and 2,000 lines.
    \item Patch Size: The maximum patch length is limited to 10,000 characters.
\end{itemize}
After filtering, we deduplicate the remaining instances against existing open-source SWE datasets. This pipeline distills an initial pool of 627k records into approximately 17k high-quality samples.

\subsection{Dataset Statistics}
\label{app:data_stats}

Table~\ref{tab:dataset_stats} presents the detailed statistics of our collected dataset, including the distribution of patch sizes and unit test breakdown. Our dataset contains 16,550 instances across 3,763 unique repositories. On average, each solution patch modifies 1.55 files and 18.76 lines of code. regarding verification, each instance includes an average of 1.98 \texttt{FAIL\_TO\_PASS} tests (verification tests) and 42.11 \texttt{PASS\_TO\_PASS} tests (regression tests).

Table~\ref{tab:dataset_comparison} compares our dataset with other standard benchmarks in the field. Our dataset provides a significantly larger scale of training instances while maintaining reasonable complexity in terms of modified files and lines.

\begin{table}[ht]
    \small
    \centering
    \caption{Detailed statistics of the collected dataset, focusing on the solution patches (Gold Patch) and evaluation unit tests.}
    \label{tab:dataset_stats}
    \setlength{\tabcolsep}{10pt}
    \renewcommand{\arraystretch}{1.2}
    \begin{tabular}{lrr}
        \toprule
        \textbf{Metric} & \textbf{Total Count} & \textbf{Avg. per Instance} \\
        \midrule
        \multicolumn{3}{l}{\textit{General Statistics}} \\
        Total Samples & 16,550 & - \\
        Unique Repositories & 3,763 & - \\
        \midrule
        \multicolumn{3}{l}{\textit{Fix Patch Statistics}} \\
        Lines Edited (Churn) & 310,544 & 18.76 \\
        Files Edited & 25,703 & 1.55 \\
        \midrule
        \multicolumn{3}{l}{\textit{Unit Test Statistics}} \\
        Fail to Pass (F2P) & 32,819 & 1.98 \\
        Pass to Pass (P2P) & 696,895 & 42.11 \\
        \bottomrule
    \end{tabular}
\end{table}

\begin{table}[ht]
    \small
    \centering
    \caption{Comparison of our dataset with related SWE datasets. We exclude execution environment details for brevity.}
    \label{tab:dataset_comparison}
    \setlength{\tabcolsep}{10pt}
    \renewcommand{\arraystretch}{1.2}
    \begin{tabular}{l c c c}
        \toprule
        \textbf{Dataset} & \textbf{\# Tasks} & \textbf{\# Repos} & \textbf{Source} \\
        \midrule
        R2E & 0.25k & 137 & Synth \\
        R2E-gym (Subset)  & 4.6k & 10 & Synth \\
        SWE-bench-extra  & 6.38k & 2k & Real \\
        SWE-bench-train & 19k & 37 & Real \\
        SWE-fixer & 115k & 856 & Real \\
        SWE-gym & 2.4k & 11 & Real \\
        SWE-smith & 50k & 128 & Both \\
        \midrule
        \textbf{SWE-World} & \textbf{16.55k} & \textbf{3,763} & \textbf{Real} \\
        \bottomrule
    \end{tabular}
\end{table}

\section{Prompt Templates}




\begin{PromptBox}[p:init_analysis_prompt]{Prompt for Initial Analysis}
\begin{lstlisting}
System Prompt:

You are a senior software engineer and bug-fixing expert.

Your task: given a bug report, human discussion, repository name, code patch, test patch,
and lists of FAIL_TO_PASS / PASS_TO_PASS tests, produce a *concise initial technical analysis*
of the problem and the fix.

**Goals of the analysis:**
- Identify the core problem / bug being fixed.
- Explain the key symptoms or incorrect behavior.
- Describe which part of the codebase (modules / functions) is conceptually responsible.
- Summarize the essence of the fix: what is changed and why it fixes the bug.
- Mention how the tests (FAIL_TO_PASS / PASS_TO_PASS) relate to the fix: what behavior they verify.
- Call out any subtle constraints / corner cases that are important.

**Style & format requirements:**
- Output MUST be plain text in English.
- Use 5-10 short bullet points (markdown "- " style).
- Each bullet should be one or two sentences, focused and technical.
- Do NOT repeat the raw diff or test lists; summarize their intent.
- Be specific but not verbose.
\end{lstlisting}

\vspace{1em}

\begin{lstlisting}
User Prompt:

### Repository
{repo}

### Problem Statement
{problem_statement}

### Human Discussion / Hints
{hints_text}

### Gold Patch (code fix)
```diff
{gold_patch}
```

### Test Patch (changes to tests)

```diff
{test_patch}
```

### FAIL_TO_PASS (should be failing before, passing after the fix)

{f2p}

### PASS_TO_PASS (should keep passing after the fix)

{p2p}

### Your Task

Produce a concise *initial technical analysis* that captures:

* what the core bug is,
* what behavior is wrong,
* what this patch is fundamentally doing to fix it,
* which areas of the code are conceptually involved,
* and how the tests validate the fix.

Remember: 5-10 bullet points, markdown - bullets, plain English, no extra commentary.
\end{lstlisting}
\end{PromptBox}

\begin{PromptBox}[p:swt_prompt]{Prompt for SWT}
\begin{lstlisting}
System Prompt:

You are an expert Python code execution simulator and a world-class software engineer. 
Your task is to predict the output of a given Python command within a specific code repository context.
Analyze all the provided information: the initial analysis, the problem description, human hints, the agent's current changes, the ideal "gold" solution, and the original content of the modified files.

Your output MUST be a single JSON object containing 'stdout', 'stderr', and 'exit_code'. Do not add any explanations or text outside of this JSON block.

### Key Information You Must Use:
1. **Initial Analysis of the Problem**: This section contains a core analysis of the issue, including the description of the error behavior, the core bug, how the issue manifests, and the intended fix. It is crucial to understanding the problem and will guide the simulation process. Use this to help you quickly identify the issue and how it should be addressed.

2. **Problem Description**: This section describes the specific issue the agent is currently working on and trying to fix. Use this to understand the exact problem the agent is attempting to resolve.

3. **Command to Simulate**: This is the command that will be executed. It contains all the information about what needs to be run, including the files to be executed and any other relevant details. Use this to simulate the execution and generate the correct output. If the "Content of Code to be Executed" is empty, the code is embedded within the command itself.

4. **Content of Code to be Executed**: This is the actual code that will be executed. Pay close attention to this content as the simulated output must strictly correspond to the code being executed. If this section is empty, the specific code to execute is provided in the "Command to Simulate" section.

5. **Agent's Current Code Modifications (Patch)**: This section highlights the changes that the agent has made to the codebase. These changes are the ones you need to analyze carefully to simulate the feedback. Focus on these modifications when generating your simulated output.

6. **Gold Standard Patch (For Your Reference)**: This is the correct solution to the issue. You should compare the agent's current changes with the gold standard solution to ensure that the simulated result is as accurate as possible. If the agent's patch is **functionally equivalent** to the gold patch (i.e., it resolves the issue in the same way), then the simulation feedback should match the expected output as defined by the gold standard.

### Your Task:
- Use all the above information to generate the most realistic and accurate simulated output.
- For commands that reproduce errors (e.g., `python reproduce_issue.py`), refer to the **Initial Analysis of the Problem** and the **Problem Description** to understand the nature of the error. Then, simulate the result based on the actual code content and the problem analysis.
- For test commands (e.g., commands that include `pytest`), carefully compare the agent's modifications with the gold standard patch. If the tests are "FAIL_TO_PASS" tests, analyze whether the agent's changes fix the issue as described in the tests. For "PASS_TO_PASS" tests, ensure that the agent's changes do not break existing functionality.
- It is important to note that the execution result must strictly follow the current code content being executed, and no fabricated test output should be added. The same test case may have multiple versions, and the test content may change across versions. Therefore, each case should be analyzed specifically based on its content.
- Your simulated output should reflect the most likely and realistic results of the command execution based on the context provided. Be precise and clear in your simulated outputs, focusing on realistic error messages, test outputs, or successful execution results.

### Format of the Output:
- **stdout**: The standard output of the command, if applicable. For example, in the case of a test run, this should reflect the results of the tests, such as which tests passed or failed.
- **stderr**: Any error messages that might be produced by the command. For example, if the command produces a syntax error or another exception, this should contain the appropriate traceback or error message.
- **exit_code**: The exit code of the command. `0` indicates success, while `1` (or any other non-zero value) indicates failure.

### Example Scenarios:
- **Successful Command Execution**: If the command executes successfully, `stdout` should contain the output generated by executing the code, and `exit_code` should be `0`. Ensure that the output corresponds directly to the expected result of the executed code.
- **Runtime Error (e.g., SyntaxError)**: If there is a runtime error, such as a syntax error, the `stderr` should contain the error traceback, and `exit_code` should be `1`.
- **Test Failures (pytest)**: If a test fails, ensure the simulated output includes the failure details (e.g., pytest output) and an appropriate `exit_code`.

Be sure to use all the context provided, including any discrepancies between the agent's modifications and the gold standard patch, to generate the most accurate simulated output.
\end{lstlisting}

\vspace{1em}

\begin{lstlisting}
User Prompt:

### 1. Initial Analysis of the Problem
{init_analysis}

### 2. Problem Description
{problem_statement}

### 3. Command to Simulate
```bash
{command_to_simulate}
```

### 4. Content of Code to be Executed
{execution_code}

### 5. Agent's Current Code Modifications (Patch)
```diff
{agent_patch}
```

### 6. Gold Standard Patch (For Your Reference)
```diff
{gold_patch}
```

### YOUR TASK
Based on all the context above, provide the simulated output for the given command. Your response must be only the JSON object, with no other text.
\end{lstlisting}
\end{PromptBox}

\begin{PromptBox}[p:swr_prompt]{Prompt for SWR}
\begin{lstlisting}
System Prompt:

You are an expert software engineering test runner and evaluator.
Your task is to simulate running a Python test command inside a code repository, and then:
1. A reasoning section enclosed within `<think>` and `</think>` tags that contains your full internal reasoning process.
2. Produce a realistic test report summarizing which tests pass or fail.
3. Decide a final reward value based on the status of specific tests.

### Key Information You Must Use:
1. **Initial Analysis of the Problem**: This section contains a core analysis of the issue, including the description of the error behavior, the core bug, how the issue manifests, and the intended fix. It is crucial to understanding the problem and will guide the simulation process. Use this to help you quickly identify the issue and how it should be addressed.

2. **Problem Description**: This section describes the specific issue the agent is currently working on and trying to fix. Use this to understand the exact problem the agent is attempting to resolve.

3. **Command to Simulate**: This is the command that will be executed. It contains all the information about what needs to be run, including the files to be executed and any other relevant details. Use this to simulate the execution and generate the correct output. If the "Content of Code to be Executed" is empty, the code is embedded within the command itself.

4. **Content of Code to be Executed**: This is the actual code that will be executed. Pay close attention to this content as the simulated output must strictly correspond to the code being executed. If this section is empty, the specific code to execute is provided in the "Command to Simulate" section.

5. **Agent's Current Code Modifications (Patch)**: This section highlights the changes that the agent has made to the codebase. These changes are the ones you need to analyze carefully to simulate the feedback. Focus on these modifications when generating your simulated output.

6. **Gold Standard Patch (For Your Reference)**: This is the correct solution to the issue. You should compare the agent's current changes with the gold standard solution to ensure that the simulated result is as accurate as possible. If the agent's patch is **functionally equivalent** to the gold patch (i.e., it resolves the issue in the same way), then the simulation feedback should match the expected output as defined by the gold standard.

7. **FAIL_TO_PASS Tests**: A list of tests that were failing before but are expected to **pass** after the correct fix. For the reward to be 1, **every** test in this list must pass.

8. **PASS_TO_PASS Tests**: A list of regression tests that were already passing and must **remain passing** after the fix. For the reward to be 1, **every** test in this list must pass.

### Your Task
Given all of the above context:
* Simulate the execution of the command under the current agent patch.
* Focus especially on the tests in FAIL_TO_PASS and PASS_TO_PASS:
  - If **all** FAIL_TO_PASS tests pass and **all** PASS_TO_PASS tests pass, then the reward **must be 1**.
  - If **any** FAIL_TO_PASS test fails or errors, the reward **must be 0**.
  - If **any** PASS_TO_PASS test fails or errors, the reward **must be 0**.

### Output Format
- **Reasoning (`<think>` block)**: First, provide your detailed internal reasoning, analysis, and step-by-step thought process enclosed within `<think>` and `</think>` tags. This section explains how you arrived at the simulated result.
- A single JSON object with the following keys:
  - **"test_report"**: A concise textual description of the simulated test results. Mention which FAIL_TO_PASS and PASS_TO_PASS tests pass or fail, and any important errors.
  - **"reward"**: An integer, either 0 or 1, following the rules above.

The final structure MUST look conceptually like this:
`<think> ...your full reasoning process here... </think>
{"test_report": "...", "reward": 0/1}`

Be sure to use all the context provided, including any discrepancies between the agent's modifications and the gold standard patch, to generate the most accurate simulated output.
\end{lstlisting}

\vspace{1em}

\begin{lstlisting}
User Prompt:

### 1. Initial Analysis of the Problem
{init_analysis}

### 2. Problem Description
{problem_statement}

### 3. Command to Simulate
```bash
{command_to_simulate}
```

### 4. Content of Code to be Executed
{execution_code}

### 5. Agent's Current Code Modifications (Patch)
```diff
{agent_patch}
```

### 6. Gold Standard Patch (For Your Reference)
```diff
{gold_patch}
```

### 7. FAIL_TO_PASS Tests (Must All Pass for reward=1)
{f2p}

### 8. PASS_TO_PASS Tests (Must All Pass for reward=1)
{p2p}

### YOUR TASK
Using all the context above, simulate running the given command, focusing on the behavior of the FAIL_TO_PASS and PASS_TO_PASS tests.

Then:
* Produce a concise test report summarizing which tests pass or fail.
* Decide the final reward:
* reward = 1 if and only if all FAIL_TO_PASS and PASS_TO_PASS tests pass.
* reward = 0 otherwise.

Your answer must include a `<think>...</think>` block containing your full reasoning process, followed immediately by a single valid JSON object with keys `"test_report"` and `"reward"`.
\end{lstlisting}
\end{PromptBox}

\begin{PromptBox}[p:cot_prompt]{Prompt for Reverse CoT}
\begin{lstlisting}
You are an expert software engineering test runner simulator and a world-class SWE bug-fix evaluator.

In this task, the TRUE test execution outcome (a JSON containing the final `test_report` and `reward`) is already provided in the input.
You MUST copy that JSON exactly at the end of your answer.

However, your reasoning MUST be written as if you do NOT know the true outcome.
You must only use the provided context (analysis, problem statement, command, code, patches, test lists, etc.) to do a forward simulation
that *could have produced* the true outcome.

The input context you receive is organized into multiple sections. Each section has a specific meaning and purpose:

1. Initial Analysis of the Problem:
   - What it is: A high-level technical analysis of the bug, including the observed error behavior, core bug, and intended fix.
   - How to use it: Use this to quickly understand what is broken, what the correct behavior should be, and what the fix is aiming to change.

2. Problem Description:
   - What it is: A more concrete description of the specific issue the agent is currently working on (bug report, task description, or failing scenario).
   - How to use it: Use this to understand the exact wrong behavior being observed and the expected correct behavior.

3. Command to Simulate:
   - What it is: The exact command that will be executed (for example, a python script invocation or a pytest command).
   - How to use it: Use this to determine the entry point, what tests or scripts are run, and how execution will proceed (including arguments, options, and test discovery).

4. Content of Code to be Executed:
   - What it is: The full content of the files and Python code that will actually be executed by the command (it may be empty if embedded in the command itself).
   - How to use it: Use this as the ground truth for runtime semantics, carefully simulating how this code will behave when run.

5. Agent's Current Code Modifications (Patch):
   - What it is: The diff of the changes that the current agent has applied to the codebase.
   - How to use it: Use this to understand exactly what logic changed, which modules/functions are affected, and how control flow or data handling is now different.

6. Gold Standard Patch (For Your Reference):
   - What it is: The correct, ideal solution (reference patch) for the same issue.
   - How to use it: Use this to compare with the agent's patch, see whether they are functionally equivalent or where they differ, and reason about possible behavioral differences.

7. FAIL_TO_PASS Tests:
   - What it is: A list of tests that were failing before but are expected to pass after a correct fix.
   - How to use it: Use these to judge whether the patch fixes the targeted bug: if any FAIL_TO_PASS test fails, the reward must be 0.

8. PASS_TO_PASS Tests:
   - What it is: A list of regression tests that were already passing and must remain passing after the fix.
   - How to use it: Use these to check for regressions: if any PASS_TO_PASS test fails, the reward must be 0.

9. True Execution Result (JSON):
   - What it is: The real outcome of running the command in a real docker environment, containing `test_report` and `reward`.
   - How to use it: You must copy this JSON exactly at the end of your answer, but you must NOT use it to guide, justify, or leak into your reasoning. Your reasoning must be written as if you have not seen this result.

---

### Your Task

A) Write a detailed forward-simulation reasoning trace that:
   - Starts from the context and code semantics.
   - Compares agent patch vs gold patch to judge whether they are functionally equivalent.
   - Predicts (as expectations) which FAIL_TO_PASS and PASS_TO_PASS tests should pass/fail and why.
      - For the PASS_TO_PASS test cases, you should carefully analyze and simulate whether the current agent patch would break any existing functionality in the repository, and use that to determine whether the tests would pass.
      - For the FAIL_TO_PASS test cases, you should carefully analyze and simulate whether the current agent patch can fix the repository's existing issue, and use that to determine whether the tests would pass.
      - This simulation should behave like a code interpreter: walk through the execution path line by line (or block by block), follow the key execution logic, and produce a complete, detailed narrative of the simulated run.
   - Based on the analysis, produce the test report for the PASS_TO_PASS and FAIL_TO_PASS tests. If all of them pass, the reward is 1; otherwise, it is 0.

B) CRITICAL CAUSALITY REQUIREMENT (NO LEAKING / NO BACKWARD EXPLANATION)
   - Your reasoning MUST NOT mention that you have been given the true result.
   - Your reasoning MUST NOT quote, paraphrase, or reference any concrete lines from the true `test_report`.
   - Do NOT write post-hoc explanations like "we see X in the output, therefore...".
   - If you mention outcomes, phrase them as expectations derived from the code, e.g.:
     "This test is expected to pass because...", "It likely fails due to...".
   - Do NOT copy any distinctive strings from the True Execution Result into your reasoning.

C) After you finish the reasoning, append the provided True Execution Result JSON EXACTLY as-is.
   - Do not reformat it, do not add keys, do not change whitespace, do not add extra fields.
   - Do not add any extra commentary after the JSON.

---

### Output Format

You must output exactly:

<sim_reasoning>
...plain-text forward simulation reasoning...
</sim_reasoning>
{PASTE_THE_PROVIDED_TRUE_EXECUTION_RESULT_JSON_HERE_EXACTLY}

Rules:
- Inside <sim_reasoning>: plain text only (no markdown headings, no code fences).
- After </sim_reasoning>: immediately paste the JSON exactly.
- Do not output anything else.
\end{lstlisting}

\vspace{1em}

\begin{lstlisting}
### 1. Initial Analysis of the Problem
{initial_analysis}

### 2. Problem Description
{problem_statement}

### 3. Command to Simulate
```bash
{command_to_simulate}
```

### 4. Content of Code to be Executed

{exec_code_block_str}

### 5. Agent's Current Code Modifications (Patch)

```diff
{agent_patch}
```

### 6. Gold Standard Patch (For Your Reference)

```diff
{gold_patch}
```

### 7. FAIL_TO_PASS Tests (Must All Pass for reward=1)

{f2p}

### 8. PASS_TO_PASS Tests (Must All Pass for reward=1)

{p2p}

### 9. True Execution Result (JSON)  [DO NOT USE THIS IN REASONING; ONLY COPY AT THE END]

{true_execution_result_json}

### YOUR TASK

Write the response in the STRICT output format required by the system prompt.
Remember:

* Reasoning must be a forward simulation based only on context.
* Do NOT reference/quote the true result in reasoning.
* Append the true JSON exactly after </sim_reasoning>.
\end{lstlisting}
\end{PromptBox}




\end{document}